\documentclass[aps,prx,twocolumn,superscriptaddress,noshowpacs]{revtex4-2}

\usepackage{times}

\usepackage{graphicx}
\usepackage{soul,color}
\usepackage{float}
\usepackage{bm}
\usepackage{braket}
\usepackage{placeins}
\usepackage{amsmath,amssymb}
\usepackage{comment}
\usepackage[dvipsnames]{xcolor}
\usepackage{mhchem}
\usepackage{url}
\usepackage[normalem]{ulem}
\usepackage{bm,nicefrac,xfrac}
\usepackage{physics}
\usepackage{xspace}
\usepackage{enumitem} 

\usepackage[normalem]{ulem}
\usepackage{xcolor}
\setstcolor{red}  

\usepackage[normalem]{ulem}

\newcommand{\fmo}{Fe$_2$Mo$_3$O$_8$\xspace}

\begin{document}

\title{Signatures of bi-altermagnetism revealed by sublattice-specific circular dichroism\\ in resonant inelastic X-ray scattering}

\author{Ganesha Channagowdra}
\affiliation{National Synchrotron Radiation Research Center, Hsinchu 300092, Taiwan}

\author{Amol Singh}
\affiliation{Department of Physics and Astrophysics, University of Delhi, New Delhi 110007, India}

\author{Hsiao-Yu Huang}
\affiliation{National Synchrotron Radiation Research Center, Hsinchu 300092, Taiwan}

\author{Mizuki Furo}
\affiliation{
Department of Physics and Electronics, Graduate School of Engineering, Osaka Metropolitan University,
1-1 Gakuen-cho, Nakaku, Sakai, Osaka 599-8531, Japan}

\author{Bin Gao}
\author{Pengcheng Dai}
\affiliation{Department of Physics and Astronomy, Rice University, Houston, 77005 Texas, USA}
\affiliation{Rice Laboratory for Emergent Magnetic materials and Smalley-Curl Institute,
Rice University, Houston, 77005 Texas, USA}

\author{Chein-Te Chen}
\affiliation{National Synchrotron Radiation Research Center, Hsinchu 300092, Taiwan}

\author{Jan Kune{\v s}}
\affiliation{Department of Condensed Matter Physics, Faculty of Science, Masaryk University, Kotl\'a\v{r}sk\'a 2, 611 37 Brno, Czechia}

\author{Atsushi Fujimori}
\affiliation{National Synchrotron Radiation Research Center, Hsinchu 300092, Taiwan}
\affiliation{Center for Quantum Science and Technology and Department of Physics, National Tsing Hua University, Hsinchu 300044, Taiwan}
\affiliation{Department of Physics, University of Tokyo, Bunkyo-Ku, Tokyo 113-0033, Japan}

\author{Sang-Wook Cheong} 
\affiliation{Keck Center for Quantum Magnetism and Department of Physics and Astronomy, Rutgers University, Piscataway, NJ 08854, USA}
\affiliation{Center for Quantum Science and Technology and Department of Physics, National Tsing Hua University, Hsinchu 300044, Taiwan}

\author{Atsushi Hariki}
\altaffiliation [Corresponding author. Email: ] {\emph{hariki@omu.ac.jp}} 
\affiliation{
Department of Physics and Electronics, Graduate School of Engineering, Osaka Metropolitan University,
1-1 Gakuen-cho, Nakaku, Sakai, Osaka 599-8531, Japan}

\author{Di-Jing Huang}
\altaffiliation [Corresponding author. Email: ] {\emph{djhuang@nsrrc.org.tw}} 
\affiliation{National Synchrotron Radiation Research Center, Hsinchu 300092, Taiwan}
\affiliation{Department of Physics, National Tsing Hua University, Hsinchu 300044, Taiwan}
\affiliation{Department of Electrophysics, National Yang Ming Chiao Tung University, Hsinchu 300093, Taiwan}

\begin{abstract}
An altermagnet is a recently identified class of magnets that exhibit a zero net magnetic moment but break symmetry under the combined operations of parity and time reversal. It typically consists of two magnetic sites of opposite spins related by rotation within the unit cell. Here, we use circular dichroism (CD)
in resonant inelastic X-ray scattering (RIXS) to reveal symmetry signatures of a distinct form of altermagnetism, namely bi-altermagnetism, in the correlated insulator Fe$_2$Mo$_3$O$_8$, which comprises two altermagnetic sublattices: one with alternating quasi-octahedral Fe environments and the other with alternating tetrahedral Fe environments. We experimentally revealed the emergence of CD in an achiral, zero-magnetization system,
thereby probing mirror-symmetry breaking associated with altermagnetic order. Notably, the CD appeared at sublattice-specific excitations of the octahedral and tetrahedral sites, indicating symmetry breaking in both altermagnetic sublattices. Calculations based on a model with the bi-altermagnetic order along the $c$ axis successfully reproduce the observed CD. Our findings support the realization of bi-altermagnetism in Fe$_2$Mo$_3$O$_8$, and showcase the use of RIXS-CD as a probe of magnetic sublattices in systems with zero net magnetization. 
\end{abstract}

\date{\today}
\maketitle

\noindent{\bf {INTRODUCTION}}

Magnetism, arising from the spin and orbital motion of electrons, is pivotal in condensed matter physics and in a wide range of technological applications. The exchange interactions between electrons determine the spin coupling, resulting in a magnetic long-range order.  Recently, a magnetic state--altermagnetism--has been proposed~\cite{Smejkal2022A,Smejkal2022B,Ahn2019,Naka2019,Smejkal2020,Hayami2019,Yuan2020}. In the standard model of altermagnetism,  up- and down-spins reside on opposite-spin crystallographic sublattices that are related to each other by a non-trivial crystallographic symmetry operation, rather than by pure inversion symmetry or translation symmetry. This unique symmetry relation enables band-structure properties typically associated with ferromagnets~\cite{Tamang2024,Chang2023,Cheong2025}, while retaining compensated magnetization in the non-relativistic limit. 
The altermagnetic symmetry gives rise to a number of interesting phenomena such as spin splitting of electronic bands~\cite{Ahn2019,Naka2019,Hayami2019,Smejkal2022B,Smejkal2020,Yuan2020,Hayami2020,Yuan2021,Mazin2021,Smejkal2022A,Liu2022,Yang2024,Yuan2024,Fernandes2024}, anomalous Hall effect~\cite{Naka2020,Smejkal2022C,Smejkal2020,Hayami2021,Mazin2021,Gonzalez2023,Naka2022}, odd magneto-optical effect~\cite{Hariki2024,Sasabe2023,Zhou2021}, chiral splitting of magnons~\cite{Smejkal2024,Liu2024}. 

\begin{figure*}[t]
\centering 
\includegraphics[width=2\columnwidth]{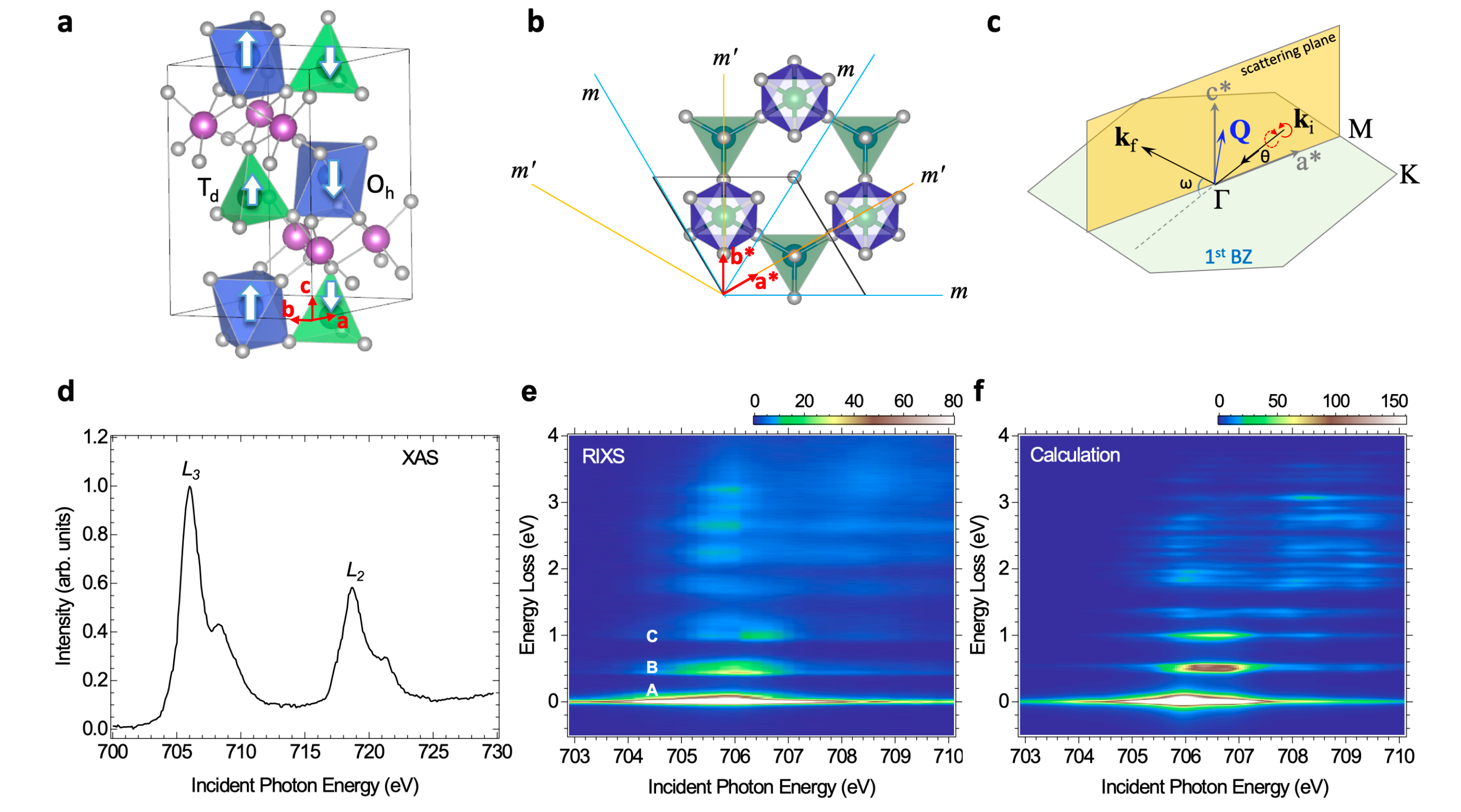}
\caption{{\bf Crystal structure and energy-dependent RIXS of \fmo}.  {\bf a}, Crystal structure of \fmo. The unit cell outlined by a black cuboid contains two quasi-octahedral ($O_h$) and two tetrahedral ($T_d$) Fe$^{2+}$ sites, whose spin moments are antiparallel, as indicated by the white arrows. {\bf b}, Illustration of the mirror planes in the magnetic point group $6'mm'$. Each of the $O_h$ and $T_d$ polyhedra in the plane $z=0$ form a triangular lattice and both together a hexagonal arrangement. Three vertical mirror planes, indicated by orange lines, correspond to the $m'$ planes, and the other three, shown in light blue, correspond to the glide planes $m$. The reciprocal lattice vectors  $\mathbf{a}^{*}$ and $\mathbf{b}^{*}$ are shown by red arrows. 
{\bf c}, Illustration of scattering geometry. The scattering plane, defined by the incident wave vector $\mathbf{k}_i$ and the scattered wave vector $\mathbf{k}_f$,  lies in the $\mathbf{a}^{*}\mathbf{c}^{*}$ plane, i.e., one of the $m'$ planes. The incident and scattering angles are $\theta$ and $\omega$,  respectively. The momentum transfer is given by $\mathbf{Q} = \mathbf{k}_f-\mathbf{k}_i$. The light blue hexagon indicates the first Brillouin zone (BZ) of the reciprocal lattice.  {\bf d}, Fe $L$-edge XAS spectrum measured in the fluorescence yield mode. The XAS is plotted without corrections for self-absorption.
{\bf e}, RIXS intensity map plotted in the plane of energy loss vs. incident photon energy. Data were measured by using right-handed circularly polarized (RCP) X-rays at 32~K for incident photon energy across the $L_3$ peak of XAS with momentum transfer $\mathbf{Q} = (0.05,0,0.65)$, i.e., $\theta = 26.26^{\circ}$ and $\omega = 70^{\circ}$. Letters A, B, and C denote the excitation energies described in the text. 
{\bf f}, Corresponding theoretical RIXS intensity map.}
\end{figure*}

Although a growing number of candidate altermagnets have been proposed through theoretical studies based on symmetry \cite{Smejkal2022B,Cheong2025}, only a limited number have been experimentally confirmed \cite{krempasky2024,Lee2024,Osumi2024,Reimers2024,Ding2024,Yang2025,Takagi2025}. This motivates the exploration of materials that preserve the essential features of altermagnetism while introducing new tunability over electronic or crystal structures. 
Here, we consider \fmo as a prototypical altermagnet composed of two altermagnetic sublattices that are not related by any crystallographic symmetry. \fmo is a multiferroic polar compound featuring two inequivalent Fe$^{2+}$ sites in quasi-octahedral ($O_h$) and tetrahedral ($T_d$) environments~\cite{Wang2015,Kurumaji2015,Reschke2020, Solovyev2019, Chang2023}, as shown in Fig.~1{\bf a}. 
It belongs to the point group $6mm$ in the paramagnetic phase and transitions to the magnetic point group $6'mm'$ with zero net magnetization in the magnetically ordered phase. In this phase, the mirror symmetries with respect to the three mirror planes are broken by the N\'eel order with the spin directions along the crystallographic $c$ axis and are restored only when combined with time reversal $\mathcal{T}$; this combined operation defines the $m'$ planes shown in Fig.~1{\bf b}. This symmetry breaking permits both symmetric and antisymmetric spin splittings \cite{Cheong2023}. As such, \fmo is predicted to be an A/S-type altermagnet \cite{Cheong2025}. Notably, from the symmetry perspective, both $O_h$ and $T_d$ sublattices in this system can individually support altermagnetic order. Therefore, we propose to call \fmo
a “bi-altermagnet," in which both sublattices independently satisfy the standard model of altermagnetism.

Identifying and further characterizing bi-altermagnetism in \fmo poses several challenges. First, the magnetic (relativistic) symmetry of the Néel state characterized by the N\'{e}el vector $\mathbf{L}{\parallel}c$ prohibits its manifestation in conventional experimental probes of time-reversal symmetry breaking associated with altermagnetism, such as the anomalous Hall effect and X-ray magnetic circular dichroism (XMCD) in absorption~\cite{Hariki2024,Gonzalez2023}. Second, both magnetic sublattices, octahedral and tetrahedral, may independently exhibit altermagnetism, complicating efforts to determine whether the concept of bi-altermagnetism is applicable to this material. For example, while direct observation of spin splitting in the band structure via angle-resolved photoemission spectroscopy could positively identify altermagnetism, it does not allow disentangling the contributions of the interpenetrating sublattices.

\begin{figure}[t]
\centering 
\includegraphics[width=1\columnwidth]{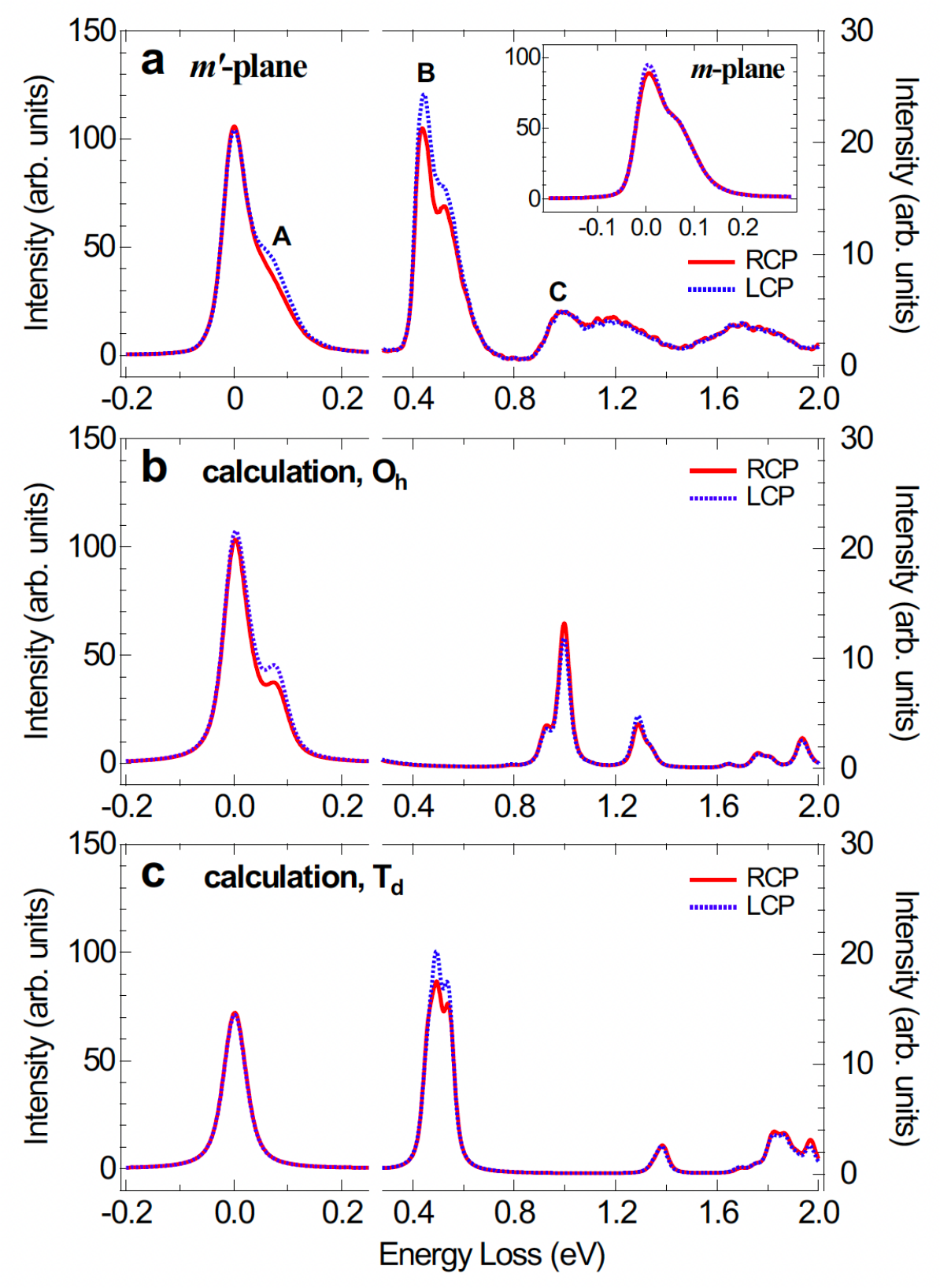}
\caption{{\bf Circular dichroism in RIXS of \fmo}. {\bf a}, RIXS spectra measured with circularly polarized X-rays and with the scattering plane in the ${\bf a}^{*}{\bf c}^{*}$ plane ($m'$ plane) at 32~K,  below the N\'eel temperature $T_{\rm N}$= 60 K. Red and blue curves, denoted by RCP and LCP, represent RIXS spectra excited with right- and left-handed circularly polarized incident X-rays, respectively. The RIXS data were recorded with momentum transfer $\mathbf{Q} = (0.05,0,0.65)$, i.e., $\theta = 26.26^{\circ}$ and $\omega = 70^{\circ}$. In the RIXS spectra plotted in {\bf a}, A indicates the low-energy spin-orbital excitations below 0.1 eV, predominantly originating from the $O_h$ site; B and C label the characteristic features of the crystal-field excitations around 0.5 eV and 1 eV at the $T_d$ and $O_h$ sites, respectively. Inset: Corresponding low-energy RIXS spectra with the scattering plane in the $ac$ plane ($m$ glide symmetry plane). {\bf b, c}, Calculations for the quasi $O_h$ and $T_d$ sites, where the contributions from the two magnetic sublattices (up- and down-spin) are summed.}
\end{figure}

In this Article, we propose and apply a method based on site-selective resonant inelastic X-ray scattering (RIXS) with circularly polarized lights to identify and characterize bi-altermagnetism in \fmo. At the Fe $L$-edge, RIXS promotes a 2$p$ core electron to the Fe 3$d$ shell, followed by the decay of a 3$d$ electron to the core level with X-ray emission. This coherent process creates a charge-neutral excitation, which may either propagate through the lattice or remain localized at the excited site. The former corresponds to collective excitations such as magnons, while the latter involves intra-atomic excitations, including spin-flip and crystal-field transitions, whose energies depend on the magnetic site and thus provide sensitivity to individual sublattices. We reveal clear circular dichroism (CD) in these sublattice-specific excitations.
The temperature dependence of the CD, together with theoretical simulations considering the bi-altermagnetic order, supports the realization of bi-altermagnetism in \fmo.

\vspace{3mm}
\noindent{\bf RESULTS AND DISCUSSION}

\vspace{3mm}
\noindent{\bf RIXS of \fmo}

We measured RIXS with the incident photon energy tuned across the Fe $L_3$ edge. Figure~1{\bf d} depicts the X-ray absorption spectrum (XAS) of \fmo at the Fe $L$ edge, while Fig.~1{\bf e} shows the RIXS intensity map plotted in the plane of energy loss versus incident photon energy. The energy levels of the $3d$ states are reversed at the two Fe$^{2+}$ sites with $O_h$ and $T_d$ symmetries. Previous density functional theory (DFT) studies~\cite{Solovyev2019,Reschke2020}, as well as our theoretical analysis presented in the Supplementary Information, show that the tetrahedral and octahedral sites have quite different crystal-field splittings, of about 0.5~eV and 1.0~eV, respectively. This difference allows us to identify site-resolved excitations in RIXS, as demonstrated in the theoretical RIXS intensity map shown in Fig.~1{\bf f} and the discussions below. Our RIXS data show that the resonance energy window of the $e \rightarrow t_2$ 
excitations ($\sim$0.5~eV) at the $T_d$ sites is significantly broader and occurs at lower incident photon energy than that of the 
$t_{2g}\rightarrow e_g$
excitations ($\sim$1~eV) at the $O_h$ sites. This site assignment is corroborated by the theoretical simulation shown in Fig.1{\bf f}. As discussed later, we also observe spin-orbital excitations below 0.1 eV (labeled A in Fig.~2{\bf a}), originating from transitions within the $t_{2g}$ manifold of the Fe$^{2+}$ ion at the $O_h$ site, with resonance energies lower than those of the crystal-field excitations.

Figure~2{\bf a} shows the RIXS spectra of \fmo with an incident photon energy tuned to the $L_3$ XAS peak (706~eV) at momentum transfer ${\bf Q}=(0.05, 0, 0.65)$ in reciprocal lattice units, with which all momentum transfers are expressed throughout the Article. The spectra exhibit characteristic features labeled A, B, and C, which are well separated in energy. These features exhibit distinct Fe-site character. Feature B is assigned to transitions between the $e$ and $t_2$ states at the $T_d$ site, while feature C corresponds to transitions between the $t_{2g}$ and $e_g$ states at the $O_h$ site. Around the energy of feature C, additional multiplet excitations ($^3T_1$, $^1A_1$) are present. These, however, have weaker RIXS intensities and originate exclusively from the octahedral site, as demonstrated by our multiplet analysis in the Supplementary Information. Furthermore, feature A represents an excitation at the octahedral site.  
We define feature A as the broad structure centered at about 70~meV that exhibits CD. It is attributed to spin-orbital excitations at the $O_h$ site, with negligible contribution from the tetrahedral site, since the spin–orbital excitation at the tetrahedral site lies well below 50~meV (see Fig.~S6(a,e)
in the Supplementary Information) and does not yield significant CD, as discussed below.

\vspace{3mm}
\noindent{\bf CD in RIXS and bi-altermagnetism}

Given the site resolution provided by the RIXS technique, we proceed to confirm the bi-altermagnetism of \fmo through the circular polarization profile and the temperature dependence of the site-specific excitations at the $T_d$ and $O_h$ sites revealed above. To this end, it is crucial to adopt an appropriate scattering geometry such that RIXS-CD directly reflects the symmetry breaking associated with bi-altermagnetic order. It should be noted that only unitary symmetries can impose constraints on the RIXS-CD intensities~\cite{Takegami2025PRL,Furo2025PRB}. This originates from the boundary conditions of the RIXS experiment: an incoming plane wave induces a resonant excitation, and the scattered wave is detected, whereas the time-reversed counterpart of this process does not contribute to the measured RIXS intensities. This asymmetry is encoded in the $i\Gamma$ term in the Kramers--Heisenberg formula. Therefore, time-reversal symmetry of the system alone cannot guarantee the vanishing of CD, as long as the resonant condition is satisfied, as time-reversal maps the above two distinct processes onto each other, rather than mapping the left- and right-circular polarizations onto each other within the actual measurement. Thus a finite CD can be present even in the paramagnetic phase,  which potentially obscures the identification of symmetry breaking at the magnetic transition. We therefore employ a geometry with the scattering plane in the $m'$ plane, as illustrated in Fig.~1{\bf c}. In this geometry, the mirror operation interchanges left- and right-circularly polarized incident X-rays, thereby enforcing the absence of CD in the paramagnetic phase, independent of the above constraint. Once this symmetry is broken by the altermagnetic order with the N\'eel vector along the $c$ axis, a finite CD can emerge below the transition temperature.

\begin{figure*}[t]
\centering 
\includegraphics[width=2\columnwidth]{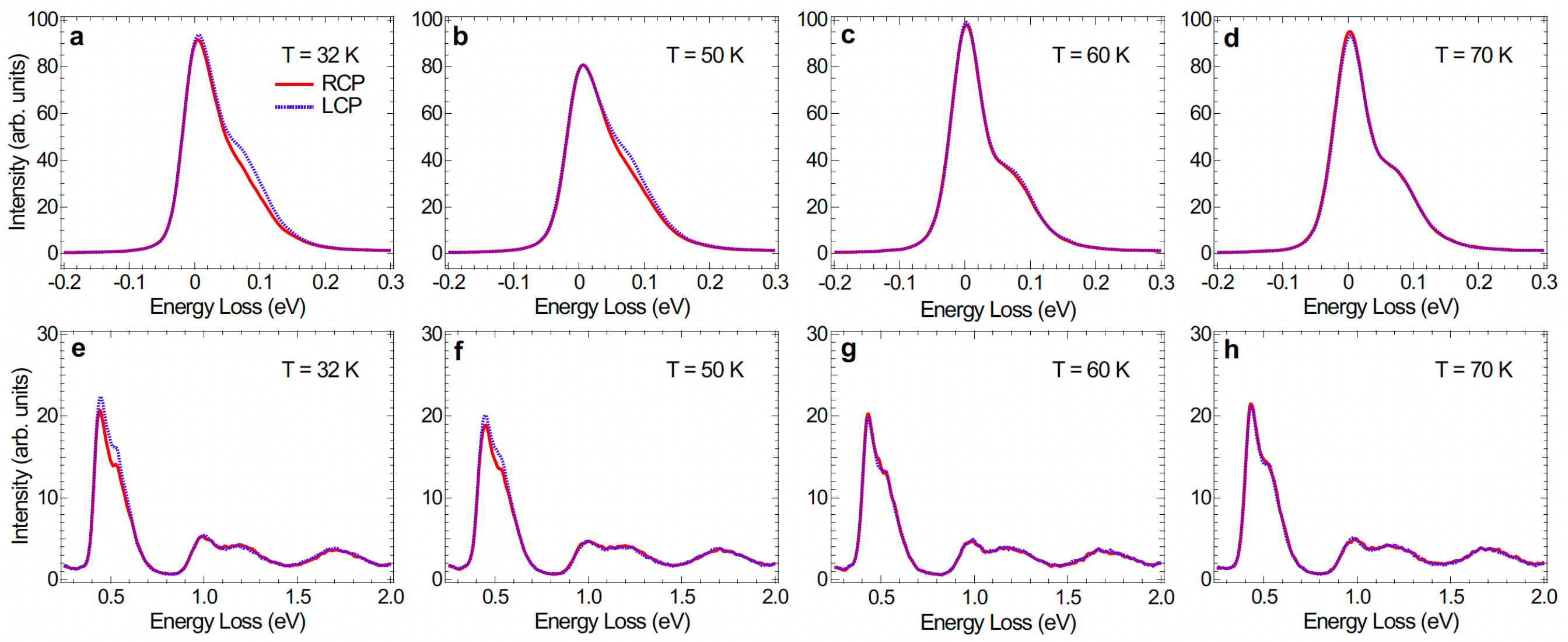}
\caption{{\bf Temperature-dependent circular dichroism (CD) in RIXS of \fmo.} {\bf a-d}, CD in RIXS arising from spin-orbital excitations near 0.1 eV. {\bf e-h}, CD in RIXS arising from $dd$ excitations. The spectra were measured at momentum transfer  $\mathbf{Q} = (0.03,0,0.66)$, i.e., $\theta = 30^{\circ}$ and $\omega = 70^{\circ}$ using circularly polarized X-rays at selected temperatures across the N\'eel temperature $T_{\rm N}$= 60 K. Red and blue curves (RCP and LCP) correspond to spectra excited with right- and left-handed circularly polarized incident X-rays, respectively. Additional RIXS-CD data at other temperatures are shown in Figs. S3 and S4 of the Supplementary Information.}
\end{figure*}

For this geometry with the $m'$ plane, we observed clear CD for both features A and B in the altermagnetic ordered phase at $T=32$~K, below the N\'eel temperature $T_{\rm N}=60$~K, as shown in Fig.~2{\bf a}. 
Figures~3{\bf a}–3{\bf h} show the temperature dependence of the RIXS CD. As the temperature increased above $T_{\rm N}$, the CD vanished within the experimental uncertainty. This temperature evolution establishes a direct link between the observed CD and the underlying altermagnetic order. 
To further verify the relation between CD in RIXS and symmetry breaking, we also measured CD for the scattering plane in the $m$ plane shown in Fig.~1{\bf b}. This mirror (glide) plane is preserved even in the altermagnetic phase, which forbids CD. Indeed, the CD for this geometry vanishes within the present experimental accuracy; see the inset of Fig.~2{\bf a} and Fig.~S2 in the Supplementary Information for detailed comparisons. This supports the conclusion that the appearance of CD in the $m'$-plane geometry reflects its symmetry breaking associated with the magnetic order.

It should be noted that the appearance of the RIXS-CD signal below $T_{\rm N}$ establishes its magnetic origin but does not, by itself, uniquely identify bi-altermagnetism, since other forms of magnetic order may also give rise to CD that disappears in the paramagnetic phase. As demonstrated by the symmetry analysis of the $m$ and $m'$ scattering geometries, the observed sublattice-specific CD follows the selection rules imposed by the magnetic point group of the bi-altermagnetic state. This symmetry-dependent response, rather than the mere onset of CD below $T_{\rm N}$, identifies the observed signal as a signature of bi-altermagnetism. For example, as discussed below, a ferrimagnetic phase with a different magnetic symmetry exhibits different selection rules for the same scattering geometries.

To further support the bi-altermagnetic origin of the observed CD and to examine their character, we performed simulations using Fe$^{2+}$ ionic models with quasi-octahedral and tetrahedral crystal fields. The parameters were derived from first-principles calculations and further refined to reproduce the energy positions of the $dd$ features in the present RIXS spectra. As each $O_h$ and $T_d$ unit contains two magnetic sites, we sum the contributions from these sites to obtain the total RIXS intensities~\cite{Hariki2024,Furo2025PRB}. Note that the two $O_h$ (and $T_d$) magnetic sites have different crystal-field Hamiltonians, mutually related by crystallographic symmetries, in contrast to conventional collinear antiferromagnetic systems where the magnetic sites share the same crystal-field Hamiltonian. To simulate the altermagnetic order (${\bf L}{\parallel}c$), we introduce a staggered Weiss (molecular) field acting on the magnetic sites within each $O_h$ and $T_d$ unit. Further details are provided in the Supplementary Information.

The simulations shown in Figs.~2{\bf b} and 2{\bf c} for both Fe sites in the altermagnetic phase reproduce the experimentally observed CD below $T_N$. The CD vanishes in the paramagnetic phase (not shown) as required by symmetry. This supports the interpretation that changes in the CD of features A and B reflect symmetry changes at both sites associated with the onset of altermagnetic order. In the models used, all excitations are confined to the excited site, and direct interactions, such as the hopping of 3$d$ electron, between the Fe sites are not taken into account. The agreement with the experimental data therefore demonstrates that the excitations exhibit a localized character and that the presence of CD directly indicates broken symmetry in the excitation spectrum at both octahedral and tetrahedral sites, supporting the realization of bi-altermagnetism in the system.

Our analysis of localized excitations using a local ionic model does not directly reveal the momentum-dependent spin splitting of the electronic band structure, a hallmark of altermagnetism. Although the magnetic symmetry of \fmo\ allows such spin splitting, its magnitude cannot be inferred from the present study and will require future experiments, for example by angle-resolved photoemission spectroscopy.

Note that feature C, which originates predominantly from crystal-field ($^5E$) and partly multiplet excitations ($^3T_1$) at the octahedral sites (see 
Fig.~S6 in the Supplementary Information), exhibits negligible CD signals in the experimental data. This behavior is not implied by symmetry, but stems from large lifetime broadening and the character of the corresponding wave functions.

Our results establish RIXS-CD as a unique spectroscopic probe to study complex altermagnetic systems and, more broadly, antiferromagnetic systems. First, CD in X-ray absorption (XMCD) of achiral systems without net magnetization usually vanishes. Indeed, the magnetic symmetry of \fmo forbids XMCD, which can be understood from the presence of three mirror (glide) planes $m$ that also prohibit the anomalous Hall effect by symmetry. While RIXS-CD has been reported in ferro- or ferrimagnetic systems~\cite{Ghosh2023,Zhang2024}, where XMCD is symmetry-allowed, CD in RIXS obeys different symmetry constraints than XMCD and can remain finite even in such a compensated magnet, as we demonstrate experimentally by explicitly measuring the emergence of the RIXS-CD signals in $dd$ features across $T_{\rm N}$; this has not been addressed in previous RIXS-CD studies on altermagnetic materials, including $\alpha$-MnTe~\cite{Takegami2025PRL,jost2025}, CrSb~\cite{Biniskos2025NC}, and $\alpha$-Fe$_2$O$_3$~\cite{Miyawaki2017}, due to their high $T_{\rm N}$, which limits access to the intrinsic absence of RIXS-CD in the normal phase. Furthermore, the interpretation of RIXS-CD in magnon excitations remains controversial~\cite{Takegami2025PRL,jost2025,Biniskos2025NC}, whereas our approach based on $dd$ excitations avoids such complexity and enables a simple model analysis directly relating the observed CD to its origin. This approach further enables
the attribution of RIXS-CD to individual sublattices, as dichroic contributions at different energies are not subject to mutual cancellation once allowed at each sublattice, thereby providing direct access to symmetry breaking at the level of the building blocks of altermagnets.

This approach can be extended to future studies of competing orders. For example, in \fmo, a ferrimagnetic state can be stabilized by applying a magnetic field or by doping~\cite{Kurumaji2015,Wang2015}, in which the magnetic moments on the sublattices within each $O_h$ and $T_d$ unit couple ferromagnetically, while the coupling between the $O_h$ and $T_d$ units is antiferromagnetic, as illustrated in Fig.~S9 of the Supplementary Information. In such a phase, the ferromagnetic alignment of the spins on each magnetic site alters the symmetry: mirror planes parallel to the $c$ axis are no longer symmetry operations of the magnetic structure. Consequently, RIXS-CD is allowed when the scattering plane coincides with the $m$ plane, a geometry that forbids CD in the altermagnetic phase. The expected amplitude of RIXS-CD is large based on theoretical simulations (see Fig.~S9). In the other geometry ($m'$ plane), CD is allowed in both altermagnetic and ferrimagnetic phases, but differs significantly in magnitude.

\vspace{3mm}
\noindent{\bf Spin-orbital excitations}

The low-energy feature $A$, where we observe a CD signal, may in principle originate from collective excitations, such as phonons and magnons, or from local spin-orbital excitations. Magnons are generally dispersive and are expected to exhibit chiral splitting~\cite{Smejkal2024,Liu2024}. Indeed, neutron scattering has reported magnon dispersion in the very low-energy range of 10--15~meV~\cite{Bao2023}. Phonon and local vibronic excitations have also been identified at intermediate energies~\cite{Reschke2020,Stanislavchuk2020}. However, the present RIXS energy resolution is insufficient to resolve such fine splittings, and the observed dispersion within the measured momentum range is small (see Fig.~S5 in the Supplementary Information). Therefore, the momentum dependence of collective excitations alone is unlikely to account for the observed CD.

In contrast, the measured CD is quantitatively reproduced by an atomic (impurity) model, indicating that it originates predominantly from localized electronic excitations at the Fe sites. This agreement further suggests that the observed CD arises primarily from the angle dependence of the polarization (dipole) operators in the RIXS scattering amplitude, rather than from the momentum dependence of the final-state wave functions.

\begin{figure*}[t]
\centering 
\includegraphics[width=2\columnwidth]{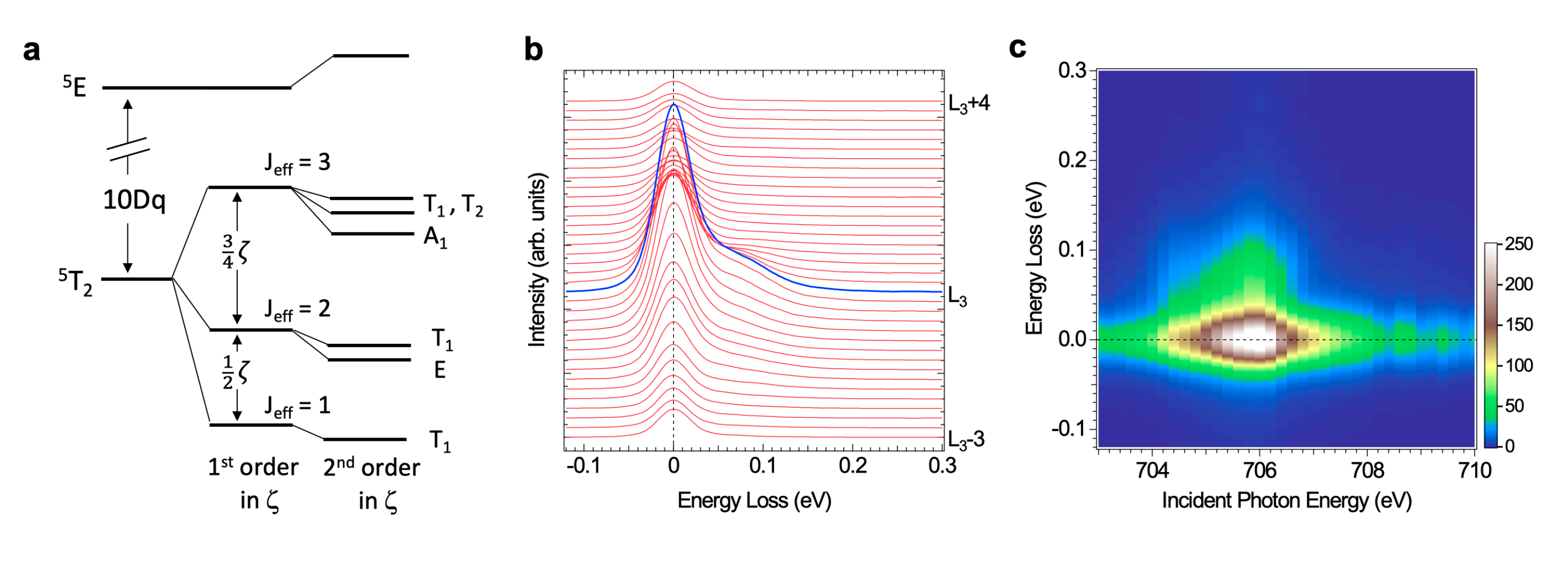}
\caption{{\bf Spin-orbital excitations of \fmo probed by RIXS}. \textbf{a}, Energy level diagram of Fe$^{2+}$ in an octahedral crystal field with spin-orbit coupling in first- and second-order approximations. For an atomic SOC constant $\zeta = 52~{\rm meV}$, atomic multiplet calculations show that the energies of the quintet substates ($E$ and $T_{1}$) are 23.7 and 25.6 meV, and those of the septet substates ($A_{1}$, $T_{1}$, and $T_{2}$) are 58.1, 62.1, and 64.8 meV. \textbf{b},~Energy-dependent RIXS measured at T = 32~K with RCP and momentum transfer $\mathbf{Q} = (0.05,0,0.65)$, i.e., $\theta = 26.26^{\circ}$ and $\omega = 70^{\circ}$. \textbf{c}, Corresponding RIXS intensity map.}
\end{figure*}

In this context, we discuss the possible spin-orbital entanglement
of the low-energy excitations and the ground state of \fmo within the ionic Fe$^{2+}$ model. The divalent Fe at the quasi-octahedral site adopts the $^5T_2$ ground-state multiplet, which hosts active spin and orbital degrees of freedom in the absence of spin-orbit coupling (SOC) within the Fe 3$d$ shell~\cite{Wei2025}. SOC splits the $^5T_2$ state into triplet, quintet, and septet states with effective angular momentum $J_{\rm eff}=1$, 2, and 3 (the lowest $J_{\rm eff}=1$ state being $T_{1g}$), as illustrated in Fig.~4{\bf a}~\cite{Lotgering1962}. For an atomic SOC constant $\zeta$, the energy separations between $J_{\rm eff}=1$ and 2, and between 2 and 3, are approximately $\zeta/2$ and $3\zeta/4$, respectively, in the first-order approximation (see the Supplementary Information). In the second-order approximation, the $J_{\rm eff}=2$ and 3 states are further split into substates. 
Figures 4{\bf b} and 4{\bf c} present the low-energy RIXS spectra and corresponding intensity map as functions of incident photon energy, covering the $^{5}T_2$ manifold. 
Assuming that $J_{\rm eff}$ provides a good description within this manifold, consistent with our simulations and previous DFT-based model~\cite{Stanislavchuk2020}, and that the effect of possible trigonal distortion is marginal in the RIXS-CD spectrum as demonstrated in the Supplementary Information, we fit the low-energy feature using two components. The resulting energy spacing follows a 2:3 ratio (see Fig.~S5), compatible with this picture.

The tetrahedral site, adopting the $^5E$ ground state, does not yield significant splitting by SOC, and its excitation remains well below 50~meV in the altermagnetically ordered phase, which has allowed us to attribute feature A predominantly to the octahedral site. While SOC leads to a singlet ground state $A_1$, the Weiss (exchange) field in the ordered phase generates a large magnetic moment, indicating Van Vleck-type magnetism at the $T_d$ site. Moreover, the ground state possesses not only spin but also orbital angular momentum, owing to the active orbital degrees of freedom, with the orbital contribution amounting to about 20\% of the spin moment in the ionic model employed (see Supplementary Information for details). The conventional concept of altermagnetism rests on the non-relativistic limit, where spin and orbital degrees of freedom are decoupled. Furthermore, in many transition-metal compounds, the 
orbital moments are often quenched in the low-energy manifold, as in established altermagnets such as $\alpha$-MnTe and CrSb. Our result highlights \fmo as a rare example among transition-metal altermagnets that retain active orbital magnetism.

\vspace{3mm}
\noindent{\bf METHODS}

\vspace{3mm}
\noindent{\bf \fmo crystal growth}

Single crystals of \ce{Fe2Mo3O8} were synthesized using a two-step method \cite{Wang2015}. The crystallinity and phase purity of the as-grown \ce{Fe2Mo3O8} crystals were initially confirmed by powder X-ray diffraction (XRD) performed on ground single crystals. The diffraction pattern matched well with the reported pattern for \ce{Fe2Mo3O8}, with no detectable impurity phases. Room-temperature refinement of the XRD data yielded a hexagonal unit cell with lattice parameters $a = 5.773(3)$\,\AA{} and $c = 10.054(3)$\,\AA{}, consistent with previous reports \cite{STROBEL1982242}. The crystal structure was indexed in the polar space group $P6_3mc$. Back-reflection Laue X-ray diffraction was used 
to orient the crystals for subsequent measurements.  These characterizations confirm that the grown samples are phase-pure and structurally well-ordered, suitable for RIXS studies. See the Supplementary Information for detailed crystal growth and characterization.

\vspace{3mm}
\noindent{\bf RIXS measurements}

We conducted Fe $L_3$-edge RIXS measurements on \fmo\ single crystals at beamline 41A of the Taiwan Photon Source \cite{SinghJSR2021}. The RIXS scattering plane was defined by the [100] and [001] directions in the reciprocal space. The incident X-ray polarization was switchable between linear and circular modes. Prior to RIXS measurements, we determined the incident X-ray energy by measuring X-ray absorption spectra recorded in the fluorescence-yield mode using a photodiode. The total energy resolution of RIXS, including contributions from both the monochromator and the spectrometer, was 45~meV with a full width at half maximum. Additional details of the RIXS measurements are given in the Supplementary Information.

\vspace{3mm}
\noindent{\bf Theoretical calculations}

We simulate the RIXS intensities using Fe$^{2+}$ ($d^6$) ionic models. 
This approach is widely used to describe localized $dd$ excitations in RIXS spectra of correlated insulators, 
including crystal-field and Coulomb multiplet excitations. 
The model Hamiltonian includes (i) valence–valence interactions within the 3$d$ shell, 
(ii) core–valence interactions between the 3$d$ and 2$p$ shells in the intermediate state of the coherent RIXS process, 
(iii) spin–orbit coupling in both the 3$d$ and 2$p$ shells, 
(iv) the crystal-field splitting of the 3$d$ levels, and 
(v) a staggered Zeeman field applied to the two magnetic sublattices within each $O_h$ and $T_d$ site to simulate the altermagnetic ordered phase. 
The crystal-field terms distinguish the $O_h$ and $T_d$ sites, whereas the other interactions are set to common values for both. 
This construction results in four atomic sites in total (up- and down-spin for both $O_h$ and $T_d$), and the site contributions are summed following Refs.~\onlinecite{Furo2025PRB,Hariki2024}.

The Hamiltonian is diagonalized numerically to obtain the full eigenvalue spectrum of the initial, intermediate, and final states, and the RIXS intensities are evaluated using the Kramers–Heisenberg formula, with an inverse lifetime broadening fixed at $\Gamma = 0.3$~eV. 
The crystal-field parameters are derived from DFT calculations using the Wien2k and Wannier90 packages~\cite{wien2k,wannier90}, based on the experimental crystal structures, yielding results consistent with previous DFT studies~\cite{Solovyev2019,Reschke2020}. 
The Slater integrals for the 2$p$--3$d$ interaction are obtained from atomic Hartree–Fock calculations and reduced to 70\% of their atomic values to account for the effects of higher configurations neglected in the atomic treatment, a well-established procedure for simulating core-level spectra at 3$d$ transition-metal edges.
The valence–valence (3$d$--3$d$) interaction is parameterized by the Hund’s coupling $J = (F_2 + F_4)/14$, which is set to 0.75~eV, a typical value for Fe-based oxides.  These parameters are optimized based on the energy positions of the $dd$ features observed in the present RIXS data. The detailed procedure is provided in the Supplementary Information.\\\\

\vspace{5mm}
\noindent\textbf{Data availability}

\noindent Source data for the figures are publicly available on Zenodo~\cite{Channagowdra2026}.

\vspace{5mm}
\noindent\textbf{Code availability}

\noindent The atomic-model calculations were performed using an in-house code that has also been used extensively in our previous studies. The code is available from the corresponding author upon reasonable request.



%

\vspace{3mm}
\noindent\textbf{Acknowledgements}

\noindent We thank the staff of the Taiwan Photon Source for technical support, and Andrei Sirenko and Hakuto Suzuki for discussions. This work was supported in part by the Taiwan National Science and Technology Council under Grant Nos. NSTC112-2112-M-213-026 (D.J.H.) and NSTC113-2112-M-007-033 (A.F.). A.S. was partially supported by ANRF (SERB) under grant number 2023/000242. We are also grateful for support from the Japan Society for the Promotion of Science under Grant No. JP22K03535. A.F. and S.W.C. acknowledge support from the Yushan Fellow Program of the Ministry of Education (MOE) of Taiwan. This work was also supported by JSPS KAKENHI Grant Numbers 25K00961, 25K07211, 23H03816, 23H03817, the 2025 Osaka Metropolitan University (OMU) Strategic Research Promotion Project (Young Researcher) (A.H.), by the project Quantum materials for applications in sustainable technologies (QM4ST), funded as project No.~CZ.02.01.01/00/22 008/0004572 by Programme Johannes Amos Commenius, call Excellent Research, and by the Ministry of Education, Youth and Sports of the Czech Republic through the e-INFRA CZ (ID:90254) (J.K.). Part of the computations in this work were performed using the facilities of the Supercomputer Center, the Institute for Solid State Physics, the University of Tokyo. The single-crystal synthesis efforts at Rice were supported by the U.S. DOE, BES under Grant No. DE-SC0012311 and DE-SC0026179 (P.D.). Part of the materials characterization efforts at Rice was supported by the Robert A. Welch Foundation Grant No. C-1839 (P.D.).

\vspace{3mm}
\noindent\textbf{Author contributions}

\noindent D.J.H. initiated the project and led the RIXS experiments. A.H. led the theoretical calculations. S.W.C. guided the symmetry discussion. G.C., A.S., H.Y.H., D.J.H., A.F., and C.T.C. conducted the RIXS measurements. B.G. and P.D. synthesized and characterized the samples. G.C, A.S., and D.J.H. analyzed the RIXS data and prepared the experimental figures. M.F., A.H., and J.K. performed the theoretical calculations. D.J.H., A.H., and A.F. wrote the manuscript with input from the other authors.

\vspace{3mm}
\noindent\textbf{Competing interests}

\noindent The authors declare no competing interests.

\end{document}



\title{{\large Supplementary Information for} \\ Signatures of bi-altermagnetism revealed by sublattice-specific circular dichroism in resonant inelastic X-ray scattering}

\author{\gc{Ganesha} Channagowdra, \gc{Amol} Singh, \gc{Hsiao-Yu} Huang, \gc{Mizuki} Furo, Bin Gao, \\ Pengcheng Dai, \gc{Chien-Te} Chen, \gc{Jan} Kune{\v s},  \gc{Atsushi} Fujimori, \\ \gc{Sang-Wook} Cheong, \gc{Atsushi} Hariki\thanks{Corresponding author: hariki@omu.ac.jp}, and \gc{Di-Jing} Huang\thanks{Corresponding author: djhuang@nsrrc.org.tw}}

\date{\normalsize\today}
\maketitle

\vspace{3cm}
\begin{center}
{\Large \bf Contents}
\end{center}
\vspace{0.5em}

\noindent
\textbf{1. Crystal growth and characterization}
\vspace{0.5em}

\noindent
\textbf{2. RIXS setup and measurements}
\vspace{0.5em}

\noindent
\textbf{3. Experimental data analysis} 
\vspace{0.5em}

\noindent
\textbf{4. RIXS simulation}

\vspace{0.5em}

\noindent
\textbf{5. Spin-Orbit Coupling in the $^5T_{2g}$ Manifold}

\newpage
\setcounter{figure}{0}
\renewcommand{\thefigure}{S\arabic{figure}}

\section{Crystal growth and characterization}

Single crystals of \ce{Fe2Mo3O8} were synthesized using a two-step method \cite{Wang2015}. First, polycrystalline \ce{Fe2Mo3O8} was prepared by a conventional solid-state reaction. Stoichiometric amounts of \ce{Fe2O3}, Mo, and \ce{MoO3} powders were thoroughly ground, pelletized, and sealed in evacuated quartz tubes. The mixture was reacted at high temperature to yield phase-pure polycrystalline material. In the second step, single crystals were grown via chemical vapor transport (CVT) using \ce{TeCl4} as the transport agent. The polycrystalline powder and a measured amount of \ce{TeCl4} were sealed together in an evacuated silica ampoule and placed in a two-zone furnace, where the source and growth zones were maintained at 958\,$^\circ$C and 853\,$^\circ$C, respectively, for about 10 days. The temperature gradient promoted directional transport of volatile species, resulting in the growth of shiny black hexagonal crystals. The resulting crystals were easily detached from the tube walls and cleaned with dilute hydrochloric acid without signs of surface oxidation.

The crystallinity and phase purity of the as-grown \ce{Fe2Mo3O8} crystals were first confirmed by powder X-ray diffraction (XRD) performed on ground single crystals. The diffraction pattern matched well with the reported pattern for \ce{Fe2Mo3O8}, with no detectable impurity phases. Room-temperature refinement of the XRD data yielded a hexagonal unit cell with lattice parameters $a = 5.773(3)$\,\AA{} and $c = 10.054(3)$\,\AA{}, consistent with previous reports \cite{STROBEL1982242}. The crystal structure was indexed in the polar space group $P6_3mc$. Back-reflection Laue X-ray diffraction was used to verify the single crystallinity and to orient the crystals for subsequent measurements.  These characterizations confirm that the grown samples are phase-pure and structurally well-ordered, suitable for further spectroscopic and scattering studies.

\begin{figure}[h]
\centering
\includegraphics[width=1\columnwidth]{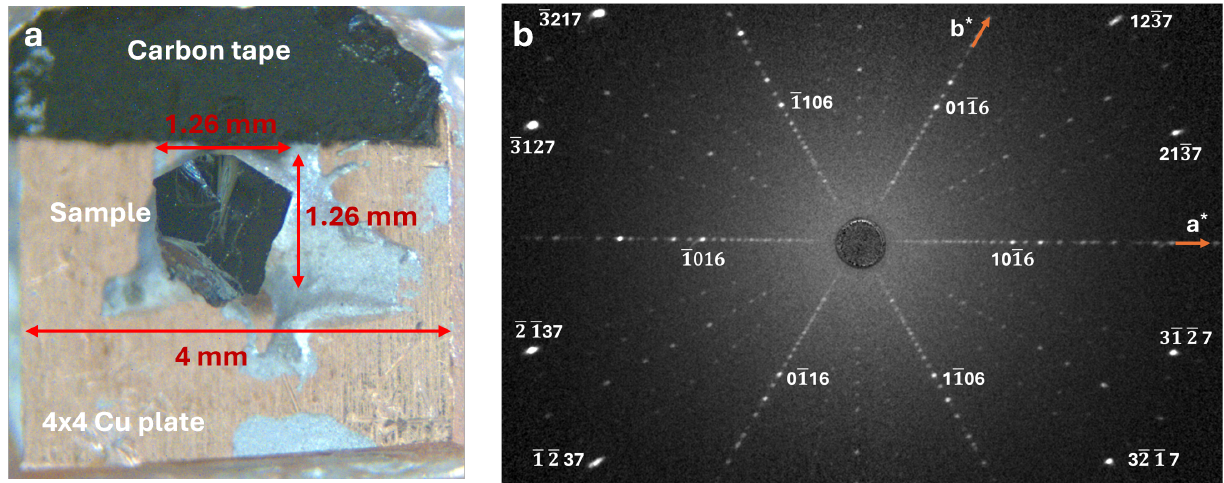}
\caption{{\bf{Sample photo and Laue pattern}. {\bf{a}}}, Photo of the \fmo sample mounted on $4mm\times4mm$ a copper plate used for RIXS measurements. {\bf{b}}, Laue pattern of the \fmo sample when the incident X-ray parallel to the reciprocal vector $c^{*}$. The orange color arrows indicate the reciprocal lattice vectors $a^{*}$ and $b^{*}$}
\end{figure}

\section{RIXS setup and measurements}

We conducted Fe $L$-edge  RIXS measurements using the AGM-AGS spectrometer of beamline 41A at Taiwan Photon Source of National Synchrotron Radiation Research Center, Taiwan \cite{SinghJSR2021}. The X-ray source is generated by an elliptically polarized undulator (EPU) in a 12-meter-long straight with a unique design. The EPU is 3.2 m long with a 48 mm period. The root-mean-square (RMS) phase error of the EPU in circular polarization mode is $\sim 3.5^\circ$, resulting in a degree of X-ray circular polarization greater than 98\%. The brilliance of the EPU tandem is designed to be greater than  $1\times10^{20}$ photons~s$^{-1}$mrad$^{-2}$mm$^{-2}$ per 0.1 $\%$ BW in the energy range of 400 to 1200 eV; the photon flux of the central cone exceeds $1\times10^{15}$ photons~s$^{-1}$. In this energy range, the calculated beam sizes are about 386 $\mu$m and 28-35 $\mu$m at the full width at half maximum (FWHM) in the horizontal and vertical directions, respectively.  

The AGM-AGS beamline is constructed based on the energy compensation principle of grating dispersion. The instrument energy resolution was 45~meV FWHM for  Fe $L$-edge RIXS measurements. The flux ratio between the left-handed circularly polarized (LCP) and the right-handed circularly polarized (RCP) X-rays is 1.1. Before the RIXS measurements, the X-ray absorption spectra (XAS) were acquired in the fluorescence-yield mode. The RIXS spectra were recorded using a bendable spherical grating and a CMOS image detector for photon-counting detection, with the polarization of the scattered X-rays not analyzed.

\section{Experimental data analysis}

\subsection{Data normalization}
The RIXS spectra were first normalized to the incident photon flux, which was inferred from the slit current, to account for fluctuations in beam intensity. Subsequently, the LCP and RCP spectra were further normalized to the integrated spectral weight in the high–energy-loss range (5–7 eV), where no spectral features are observed. This procedure ensures a reliable comparison between the two polarization channels and eliminates artificial intensity offsets. Note that, because the EPU delivers X-rays with a high degree of circular polarization, well within the fluctuations of the measured CD. Accordingly, the measured RIXS-CD spectra are presented without normalization by the incident circular polarization.    

\subsection{Curve fitting}

In principle, the measured low-energy RIXS spectra contain contributions from electromagnons, magnons, phonons, and spin–orbital excitations. However, electromagnon (4–10 meV) and magnon (11 and 14 meV) energies lie well below the instrumental resolution (45 meV FWHM) and are thus buried within the elastic peak, rendering them unresolvable. 



To identify the energy position and momentum-dependent dispersion of low-energy excitations of \fmo measured at $\mathbf{Q}= (0, 0, L)$ direction, we performed a least-squares curve fitting of the RIXS spectra by using three components: one Voigt profile for the elastic peak and two Lorentzian profiles for the low-energy excitations, as shown in Figure {\ref{fig_dispersion}. In addition, a linear background was incorporated to ensure an accurate fit. The quasi-elastic contribution is constrained by the independently measured instrumental resolution from the elastic peak of a carbon tape both before and after each RIXS spectrum measurement. 

Our spectral analysis should therefore be regarded as a constrained minimal decomposition of the measured line shape, rather than as evidence that the lowest spin-orbital component is fully resolved from the quasi-elastic line. Specifically, the spectra are fitted using a minimal $(2+1)$-component model. The lowest spin-orbital component is required to reproduce the asymmetric spectral weight on the positive-energy-loss side of the elastic line.

\begin{figure}[h]
\centering
\includegraphics[width=0.80\columnwidth]{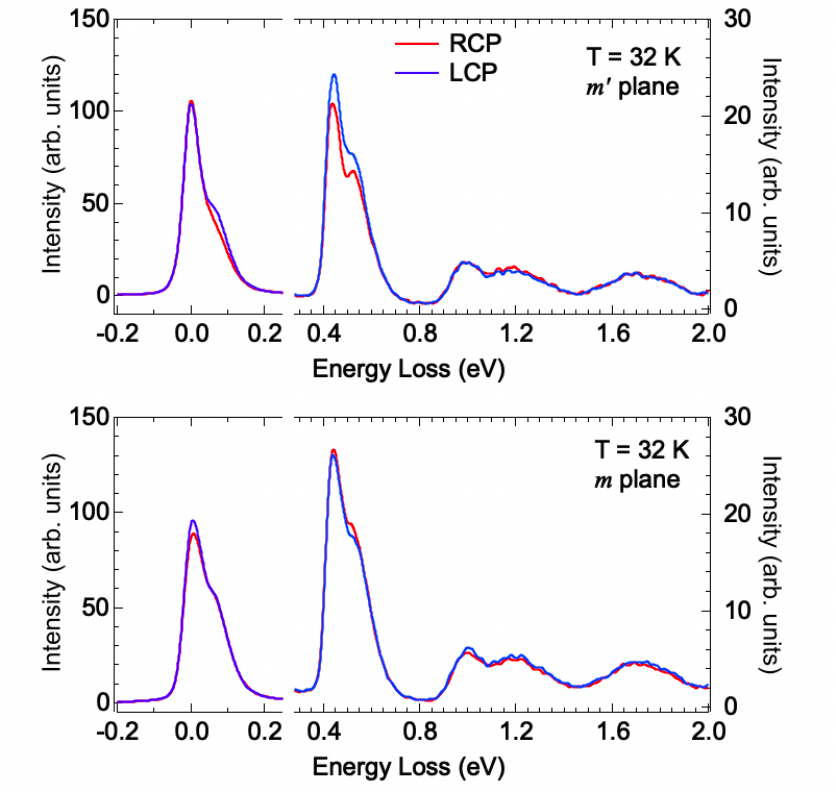}
\caption{{\bf Comparison of CD in RIXS of \fmo}. Upper Panel: RIXS spectra measured with circularly polarized X-rays and with the scattering plane in the $a^{*}c^{*}$ plane ($m'$ symmetry plane) at 32~K,  below the N\'eel temperature $T_{\rm N}$= 60 K. Red and blue curves, denoted by RCP and LCP, represent RIXS spectra excited with right- and left-handed circularly polarized incident X-rays, respectively. The RIXS data were recorded with momentum transfer $\mathbf{Q} = (0.05,0,0.65)$ i.e., $\theta = 26.26^{\circ}$ and $\omega = 70^{\circ}$. Lower Panel: Corresponding RIXS spectra measured with the scattering plane in the $ac$ plane ($m$ symmetry plane) at 32~K.} 
\label{fig_m_vs_m'}
\end{figure}

\begin{figure}[h]
\centering
\includegraphics[width=0.80\columnwidth]{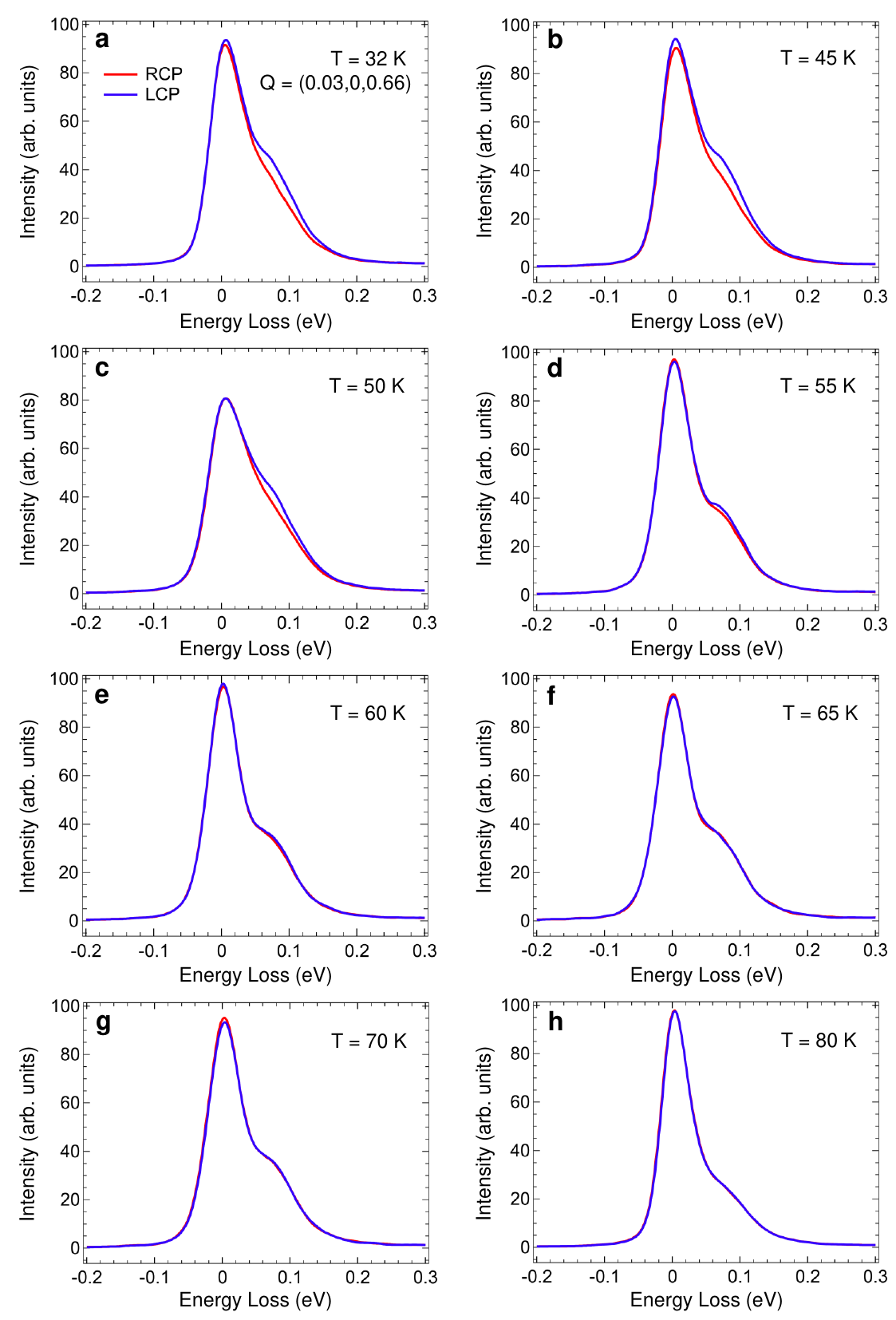}
\caption{{\bf Temperature-dependent CD}. {\bf a-h}, Temperature-dependent CD in the low-energy spin-orbital excitations of \fmo measured across the $T_N$ with circularly polarized X-rays at $\mathbf{Q}= (0.03, 0, 0.66)$ i.e., $\theta = 30^{\circ}$ and $\omega = 70^{\circ}$. Red and blue curves, denoted by RCP and LCP, represent RIXS spectra excited with right- and left-circularly polarized incident X-rays, respectively.}
\label{fig_S1}
\end{figure}

\begin{figure}[h]
\centering
\includegraphics[width=0.80\columnwidth]{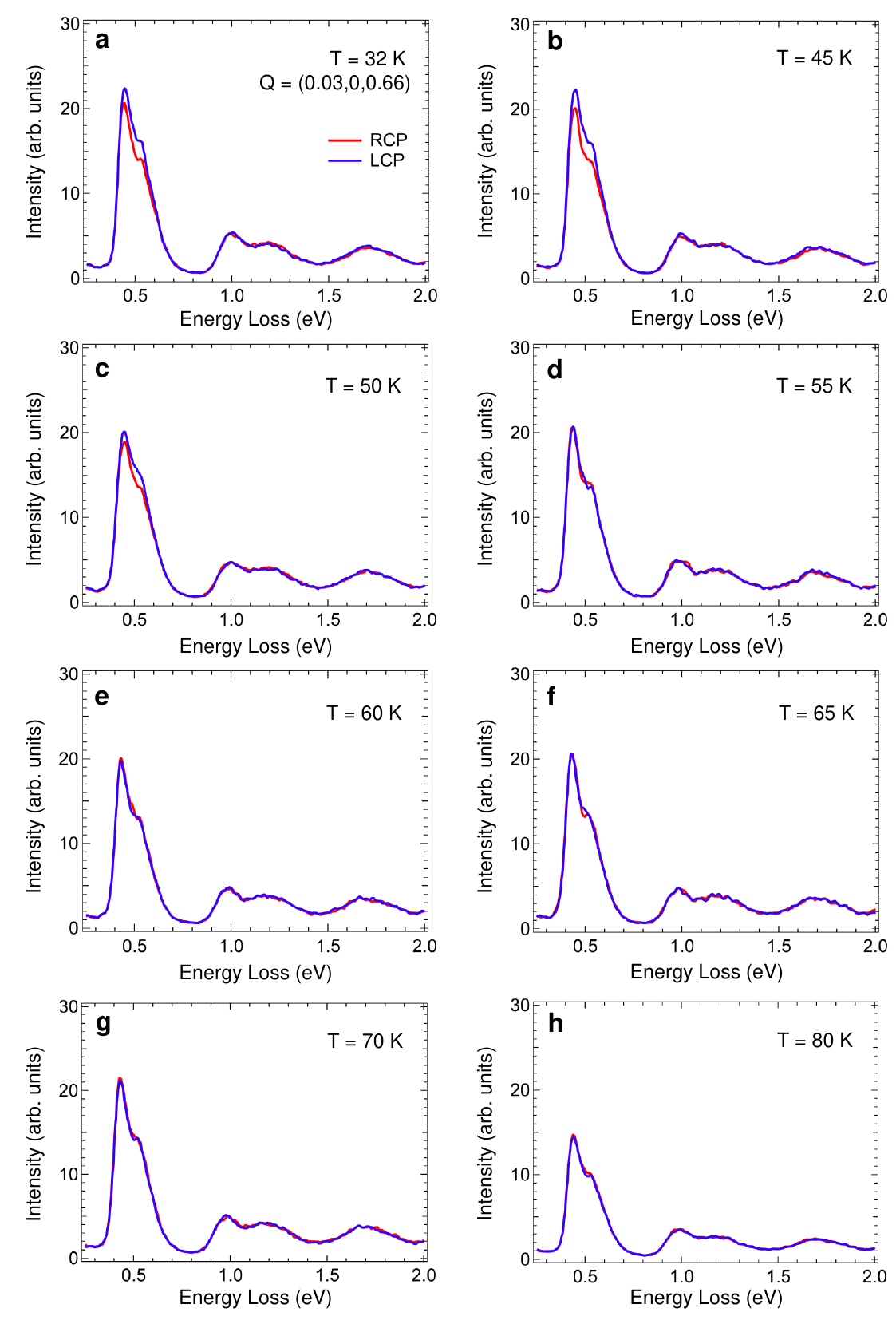}
\caption{{\bf{Temperature-dependent CD}. {\bf{a-h,}}} Temperature-dependent CD in the crystal-field excitations of \fmo measured across the $T_N$ with circularly polarized X-rays at $\mathbf{Q}= (0.03, 0, 0.66)$ i.e., $\theta = 30^{\circ}$ and $\omega = 70^{\circ}$. Red and blue curves, denoted by RCP and LCP, represent RIXS spectra excited with right- and left-circularly polarized incident X-rays, respectively.}
\label{fig_S2}
\end{figure}

\begin{figure}[h]
\centering
\includegraphics[width=0.8\columnwidth]{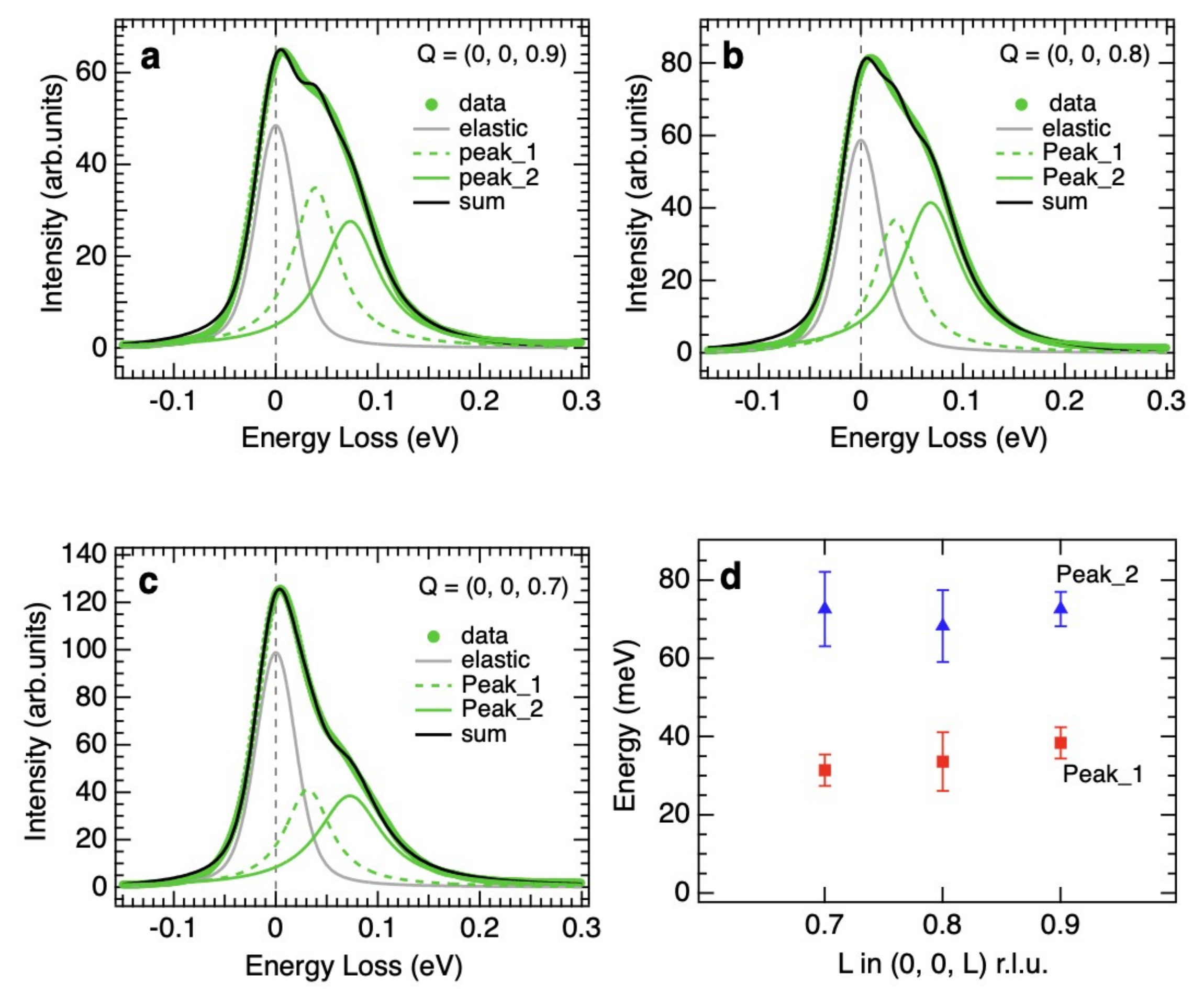}
\caption{{\bf{Fitting results for the Q-dependent RIXS}. {\bf{a-c,}}} Momentum-dependent dispersion of spin-orbital excitations of \fmo measured at $\mathbf{Q}= (0, 0, L)$ with right-circular polarized X-ray. The green solid circle indicates the experimental data, and the black solid line represents the sum of the fitting peaks. The fitted model consists of a Voigt profile (gray solid line) for the elastic peak and Lorentzian profiles for the spin–orbital excitations, represented by the green dotted and solid lines for peak~1 and peak~2, respectively. {\bf d,} Dispersion of peak~1 (red square) and peak~2 (blue triangle) extracted from the fitting results.}
\label{fig_dispersion}
\end{figure}


\section{RIXS simulation}

Here we describe our RIXS simulation of \fmo using an ionic model. We first construct two ionic models representing the quasi-$O_{h}$ and $T_{d}$ sites, starting from density functional theory calculations. The derived parameters are then further optimized under three criteria ($C1$, $C2$, $C3$) imposed by the present RIXS data shown in Fig.~\ref{fig_theo_s1}(a). Based on these optimized ionic models, we proceed with the simulation of the RIXS-CD of \fmo.

\begin{enumerate}[label=$C\arabic*$.]
  \item 
    Both Fe sites adopt a divalent $d^6$ configuration with well-localized $d$ electrons. Therefore, we use the same interaction parameters for the 3$d$--3$d$ (valence–valence) and 2$p$--3$d$ (core–valence) Coulomb multiplet interactions at the two Fe sites. For the core–valence interaction, the Slater integrals are computed using atomic Hartree–Fock calculations, and the values are reduced to 70\% of the bare values to account for the effect of higher configurations, which is a well-established empirical treatment for simulating core-level spectra at 3$d$ transition metal edges~\cite{Hariki2017,Matsubara2005,Sugar1972,Tanaka1992,Groot1990}. The 3$d$--3$d$ interaction is parameterized by the Hund’s parameter $J_{\rm H} = (F_2 + F_4)/14$, where $F_2$ and $F_4$ are Slater integrals. We set $J_{\rm H} = 0.75$~eV, which is a typical value for Fe-based oxides, and its validity is confirmed by the position of the $dd$ features, as explained in criteria~3. The two Fe sites are distinguished by the crystal-field terms in the corresponding ionic Hamiltonian, which are determined straightforwardly from the experimental criteria~2 ($T_{d}$ site) and~3 ($O_{h}$ site) described below.
  \item 
    The $T_{d}$ site exhibits a distinct $dd$ excitation (feature $B$) around 0.4--0.5~eV, Fig.~\ref{fig_theo_s1}(a), corresponding to the $^5T$ excitation (in the absence of the spin-orbit coupling). The excitation energy from the ground state ($^5E$) to this excited state is determined by the crystal-field splitting ${\rm CF}_{T_d}$ between the $t$ and $e$ orbitals in the $T_{d}$ crystal environment, which allows us to estimate its reasonable value as ${\rm CF}^{\rm opt}_{T_d}=0.49$~eV, see the energy diagram in Fig.~\ref{fig_theo_s1}(b). In the experimental data, feature $B$ exhibits fine structure arising from the spin–orbit coupling ($\xi_{3d}$) within the Fe 3$d$ shell. As shown in Fig.~\ref{fig_theo_s1}(c), $\xi^{\rm opt}_{3d} = 65$~meV yields a splitting reasonably consistent with the experimental data. We therefore adopt this value at both the $O_{h}$ and $T_{d}$ sites in simulating the RIXS-CD. This value is slightly larger than the one derived from atomic calculations ($\xi_{3d}^{\rm atom} = 52$~meV). The discrepancy may be related to hybridization with ligand or Mo sites, which is not explicitly treated within the present ionic models. Note that we have checked that the presence or absence of CD is irrelevant with varying the spin–orbit coupling constant within this scale.
  \item 
    The $O_{h}$ site exhibits multiple $dd$ features, labeled $C$, with a pronounced peak at approximately 1.0~eV and an additional feature around 1.2~eV, indicated as $C'$ in Fig.~\ref{fig_theo_s1}(a). In Fig.~\ref{fig_theo_s1}(d), we show the calculated energy diagram as a function of the crystal-field splitting ${\rm CF}_{O_h}$between $t_{2g}$ and $e_g$, where spin–orbit coupling within the Fe 3$d$ shell is neglected for clarity, as it is not essential for the optimization of the $O_h$ site discussed below. As seen in the simulated energy profile, the $^5E_g$, $^3T_{1g}$, and $^1A_1$ states are nearly degenerate at 1.0~eV. For the high-spin ground state ($^5T_2$), however, the $^5E$ excitation exhibits a stronger RIXS intensity owing to a spin-conserving process, and thus serves as the dominant contribution to the 1.0~eV $C$ feature, compared to the intermediate-spin excitation $^3T_{2}$. The RIXS intensity of $^1A_1$ is negligibly weak, consistent with $L_3$-edge RIXS studies on similar $d^6$ systems~\cite{Wang2018,Hariki2020}. The higher intermediate-spin excitation $^3T_{2}$ gives rise to the feature around 1.2~eV. We stress that the excitation energy of the $^5E$ state is exclusively determined by the crystal-field splitting within the model, irrespective of the $J_{\rm H}$ value, since it has the same high-spin character as the ground state. This allows us to set the optimal crystal-field splitting ${\rm CF}^{\rm opt}_{O_h}$ to 0.95~eV. By contrast, the intermediate-spin states ($^3T_{1}$, $^3T_{2}$) depend on $J_{\rm H}$, and only a narrow range around $J_{\rm H} = 0.75$~eV yields agreement between their energies and the experimental data (see Fig.~\ref{fig_theo_s1}e). Simultaneously, the simulated spectra adopting ${\rm CF}^{\rm opt}_{O_h}$ and $J_{\rm H} = 0.75$~eV reproduce additional features in the range 1.6–2.0~eV. In the RIXS-CD simulations below, we include spin–orbit coupling with $\xi^{\rm opt}_{3d} = 65$~meV in the derived ionic model at the $O_h$ site.    
 \end{enumerate}

\begin{figure}[h]
\centering
\includegraphics[width=0.99\columnwidth]{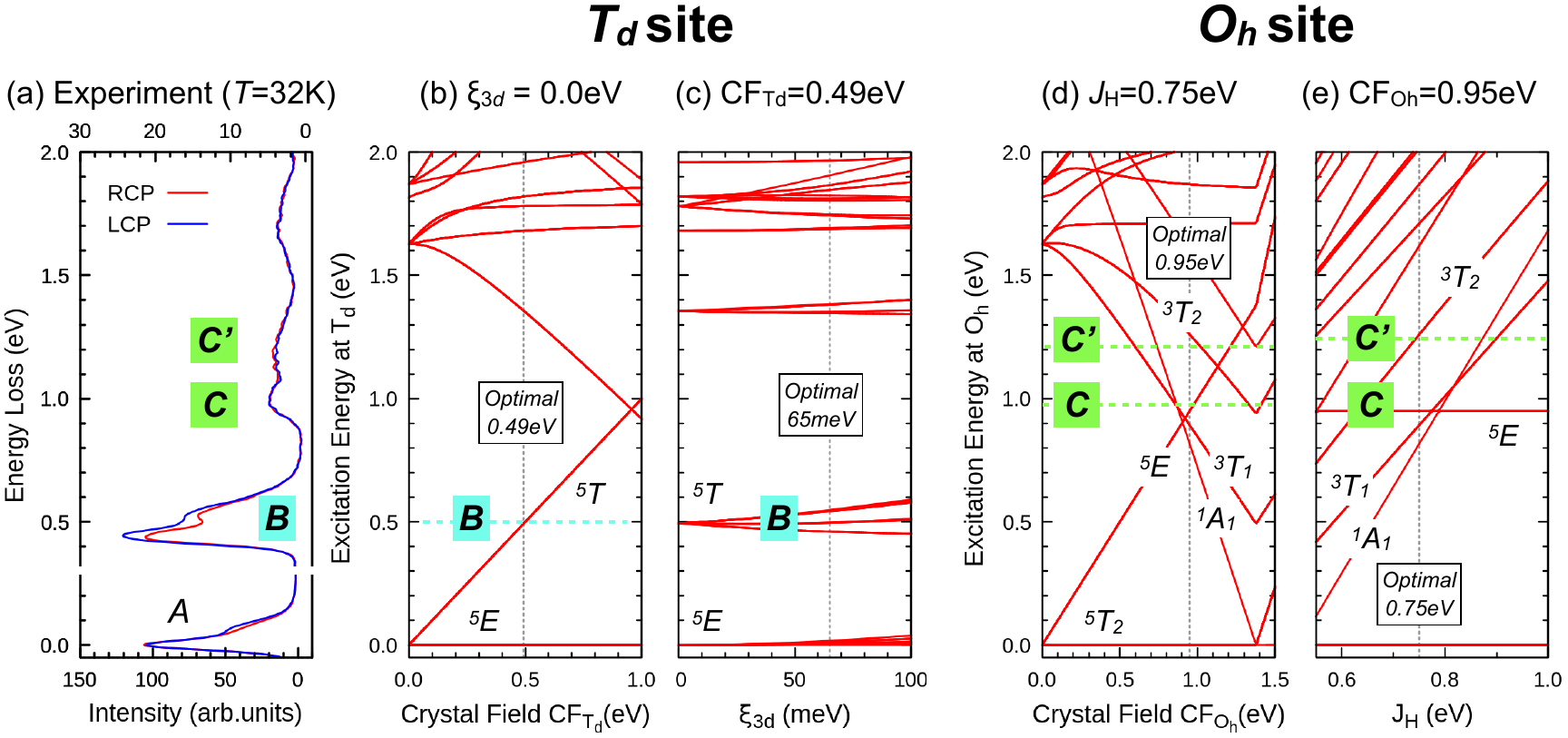}
\caption{(a) Experimental Fe $L_3$-edge RIXS spectra measured with circularly polarized X-rays, reproduced from Fig.~2{\bf a} in the main text. (b, c) Energy diagrams computed using the atomic model for the $T_{d}$ site: (b) as a function of the crystal-field splitting ${\rm CF}_{T_d}$ (with spin–orbit coupling set to zero), and (c) as a function of the spin–orbit coupling $\xi_{\rm soc}$ for the optimal crystal-field splitting (0.49~eV). (d, e) Energy diagrams computed for the $O_{h}$ site: (d) as a function of the crystal-field splitting ${\rm CF}_{O_h}$ and (e) as a function of the Hund’s coupling $J_{\rm H}$ for the optimal crystal-field splitting (0.95~eV), where spin–orbit coupling is neglected for clarity. In each panel, vertical bars indicate the optimal value reproducing the characteristic peak positions observed in the experimental RIXS data.}
\label{fig_theo_s1}
\end{figure}

\begin{figure}[h]
\centering
\includegraphics[width=0.99\columnwidth]{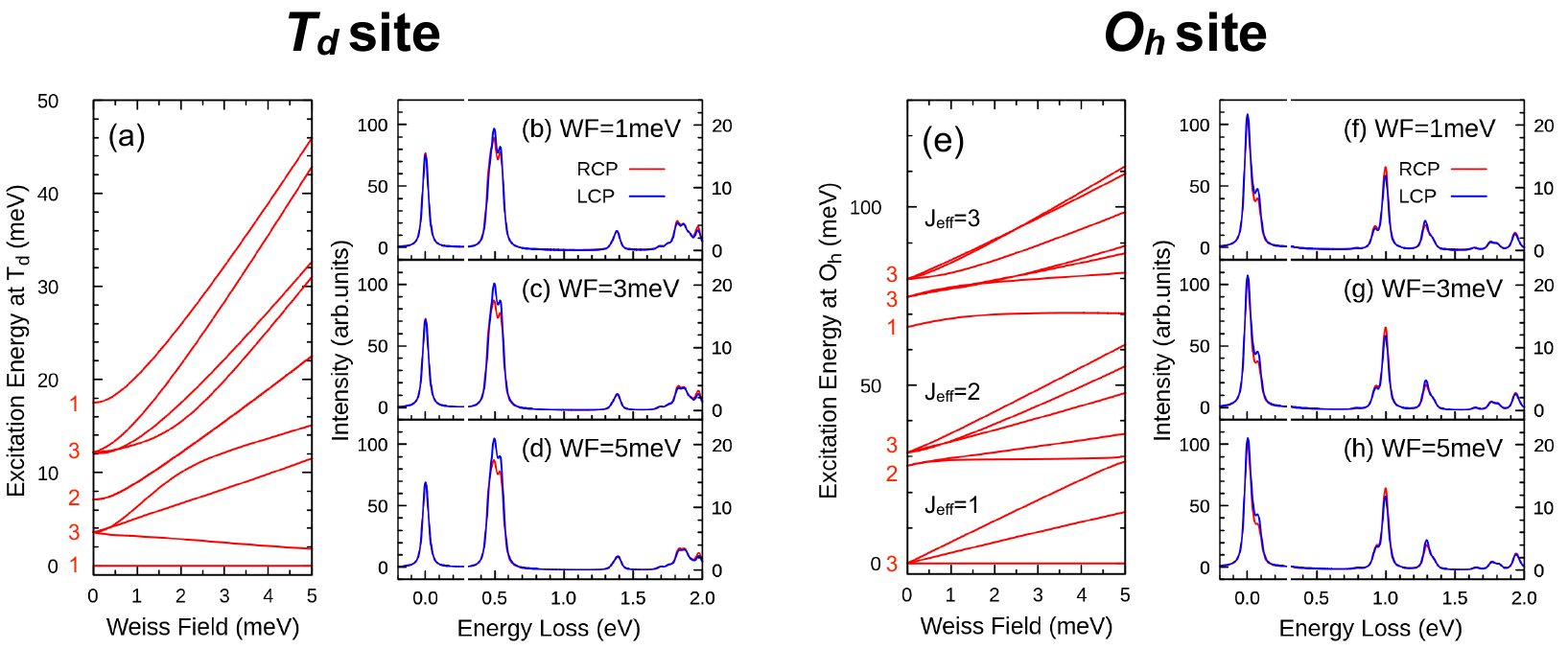}
\caption{\gc{(a, e)} Evolution of the excitation energies with the altermagnetic Weiss (molecular) field at the $T_d$ (left) and $O_h$ (right) sites. The multiplicities of the excited states in the absence of the Weiss field are indicated in the panels. (b–d, f–h) Simulated RIXS-CD spectra for three selected amplitudes of the Weiss field (1, 3, and 5~meV) are shown for each Fe site.}
\label{fig_theo_s2}
\end{figure}


We note that the above optimization of the model relies only on the RIXS spectra, and the CD profile is not used in this process. We now proceed with the analysis of the RIXS-CD based on the optimized ionic model. To this end, we introduce an effective Weiss (exchange) field to represent the altermagnetic order in the lattice. As discussed in the main text, the ground-state manifold in the non-relativistic limit (i.e., without spin–orbit coupling) is $^5T_2$ at the $O_h$ site and $^5E$ at the $T_d$ site. Given the degrees of freedom in both spin and orbital angular momentum in the ground-state manifold, the effective field may in principle take a complex form, depending on both. In the present work, however, we do not attempt to derive the detailed form of the Weiss field, but instead restrict ourselves to the simplest form, $H_{\rm Weiss} = -{\bf S}\cdot{\bf B}_{\rm eff}$, assuming that it acts only in the spin space and that the field direction is parallel to the N\'eel vector. This simplified form reproduces the experimental data reasonably well, as shown in the main text and in the supplemental calculations below. Each $O_h$ and $T_d$ unit consists of two magnetic sublattices (i.e., four ion sites in total in the simulation). A staggered Weiss field is applied to the sublattices to simulate the antiferromagnetic arrangement of the magnetic moments~\cite{Furo2025PRB,Hariki2024}. The total contributions from the $O_h$ and $T_d$ sites are then obtained by summing over the two magnetic sublattices within each unit.

Figures~\ref{fig_theo_s2}(a)(e) show the evolution of the low-energy spectrum with the Weiss field at the $T_d$ and $O_h$ sites, respectively. In Fig.~\ref{fig_theo_s2}(e), at the $O_h$ site, spin–orbit coupling lifts the degeneracy of the $^5T_2$ ground state, splitting it into $J_{\rm eff}=1, 2, 3$, as discussed in the main text. The Weiss field introduces a Zeeman effect in the $J_{\rm eff}=1$ state, leading to a magnetic ground state. When a small distortion at the $O_h$ site is taken into account, the spectrum is slightly modified, but the RIXS-CD remains unchanged, as we demonstrate below. In Fig.~\ref{fig_theo_s2}(a), at the $T_d$ site, spin–orbit coupling stabilizes a singlet ground state out of the $^5E$ manifold. However, when the Weiss field is turned on, both a sizable spin and an orbital angular momentum emerge in the ground state. For example, Weiss-field amplitudes of 1, 3, and 5~meV result in $(S_c, L_c) = (1.29, 0.30)$, $(1.67, 0.35)$, and $(1.80, 0.35)$, respectively, where $c$ denotes the crystallographic $c$ direction. The sign of the moments is reversed between the magnetic sublattices, leading to compensation of the total magnetic moment. This behavior indicates that the magnetism at the $T_d$ site is of the Van Vleck type. We have calculated the RIXS-CD from the ground state for selected amplitudes of the Weiss field. At around 3–5~meV, which appears to be a reasonable value for the studied system, we observe CD at features $A$ and $B$. As explained in the main text, the former and latter originate from the $O_h$- and $T_d$-site contributions, respectively.


\begin{figure}[h]
\centering
\includegraphics[width=0.99\columnwidth]{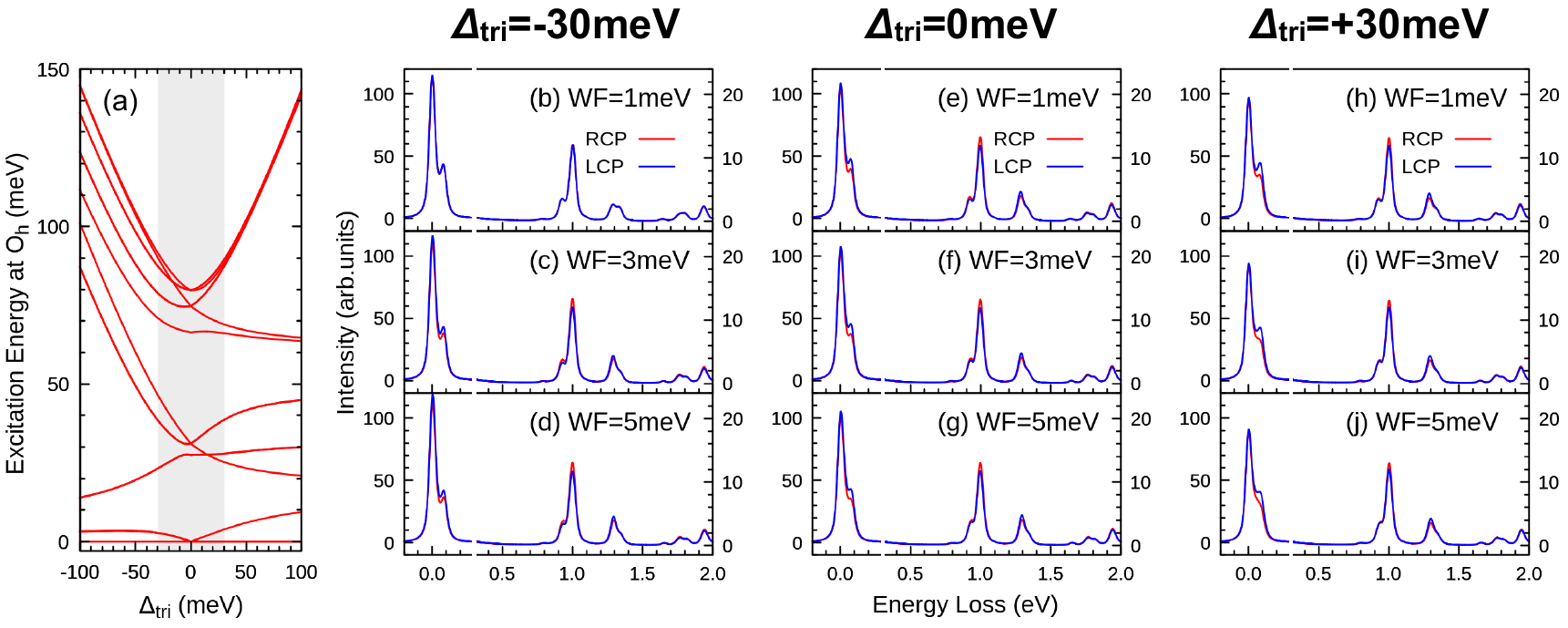}
\caption{(a) Evolution of the excitation energies with the trigonal distortion $\Delta_{\rm tri}$ at the Fe $O_h$ site. \gc{(b-j)} \gc{S}imulated RIXS-CD spectra for three selected Weiss-field amplitudes (1, 3, and 5~meV) for $\Delta_{\rm tri}=-30, 0, 30$~meV are shown for comparison.}
\label{fig_theo_s3}
\end{figure}

Finally, we comment on the effect of a distortion of the surrounding oxygens. At both the $O_h$ and $T_d$ sites, a small trigonal-like distortion splits the $t$ level into a singlet ($a_1$) and a doublet ($e$), with the splitting denoted as $\Delta_{\rm tri}$ hereafter. As shown in Refs.~\cite{Solovyev2019,Reschke2020} and confirmed by our DFT simulations, the splitting at the $T_d$ site is tiny. Furthermore the ground-state manifold at the $T_d$ site is $^5E$, which is essentially unaffected to the $\Delta_{\rm tri}$ splitting. This justifies neglecting the splitting in our analysis. For the $O_h$ site, however, the $^5T$ ground-state degeneracy should be lifted by $\Delta_{\rm tri}$, while both the magnitude and the sign of $\Delta_{\rm tri}$ appear to be difficult to determine~\cite{Solovyev2019,Reschke2020}. 

To provide support that the ambiguity in the distortion at the $O_h$ site does not affect our analysis of RIXS-CD, we calculated the low-energy excitation spectrum in the range $\Delta_{\rm tri} = -100$ to $100$~meV. Depending on the sign of $\Delta_{\rm tri}$, a singlet (negative $\Delta_{\rm tri}$) or doublet (positive $\Delta_{\rm tri}$) ground state is realized. The present RIXS experiment, in Fig.~\ref{fig_dispersion}(d), clarifies at least two spectral features centered around 35~meV and 70~meV, with minimal momentum dependence. The cases $\Delta_{\rm tri} = -30$~meV and $30$~meV appear consistent with the experimental constraints, and the magnitude is also consistent with previous theoretical estimates. In Fig.~\ref{fig_theo_s3}, we present the calculated RIXS-CD spectra for $\Delta_{\rm tri} = -30$, 0, and 30~meV for selected amplitudes of the Weiss field (WF). The evolution of the excitation spectrum with the Weiss field is shown in Fig.~\ref{fig_theo_s3}(a). We find that all cases yield very similar RIXS-CD spectra, with a peak at 70~meV, in agreement with the present RIXS-CD experiment. Therefore, we conclude that the effect of the small distortion is not relevant for the main discussion of the present work.

\begin{figure}[h]
\centering
\includegraphics[width=0.99\columnwidth]{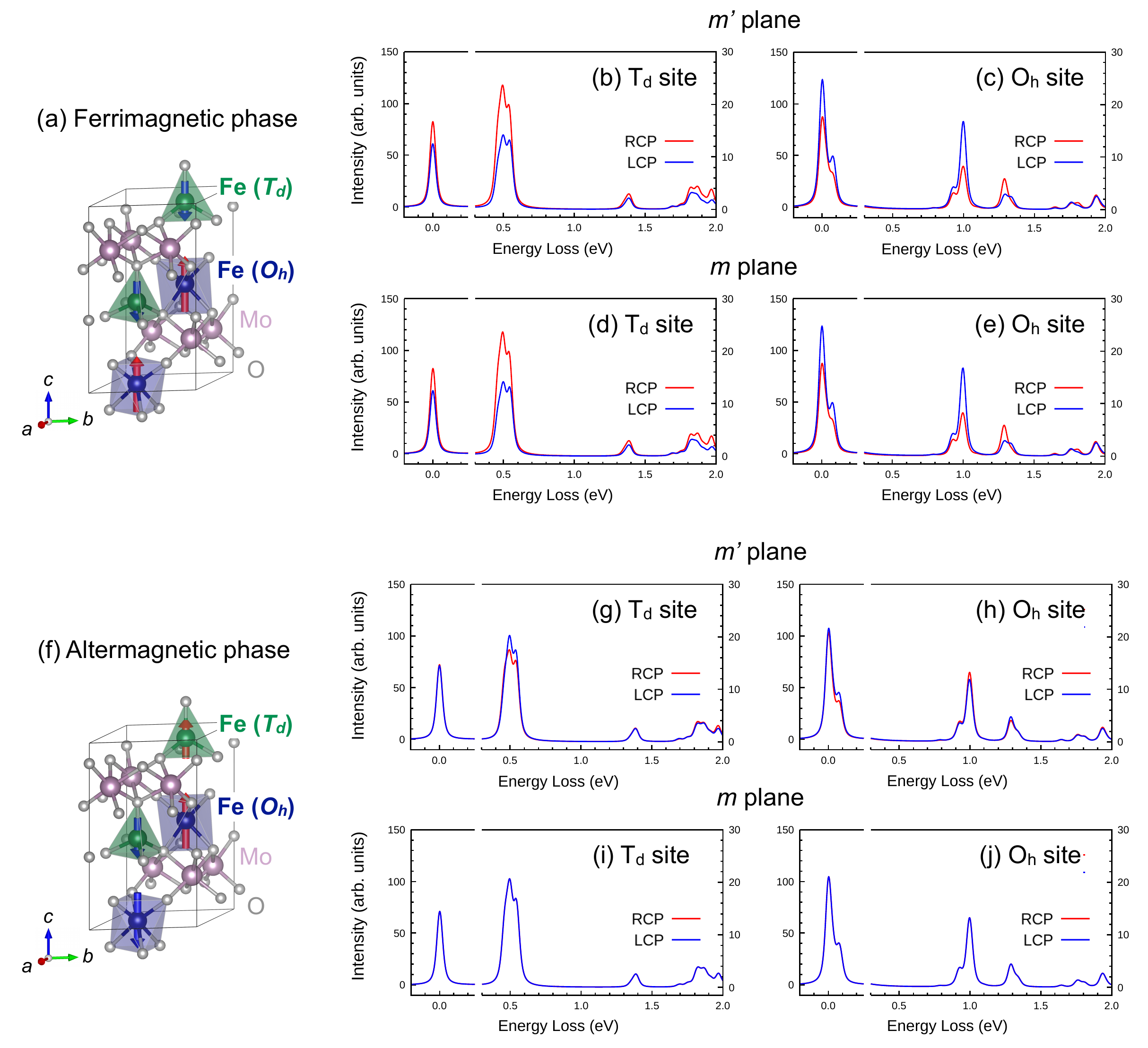}
\caption{(a, f) Schematic illustration of the ferrimagnetic and bi-altermagnetic structures. The magnetic moments are indicated by arrows. (b–e, \gc{g-j}) Simulated RIXS-CD spectra for the ferrimagnetic (b–e) and bi-altermagnetic (\gc{g-j}) phases with the $m'$ and $m$ scattering planes at the $T_d$ and $O_h$ sites. The sublattice contributions in each $T_d$ and $O_h$ unit are summed.}
\label{fig_theo_ferri_alt}
\end{figure}

In \fmo, it is known that a ferrimagnetic state can be stabilized by applying a magnetic field or by doping, where the magnetic moments on the sublattices within each $O_h$ and $T_d$ unit couple ferromagnetically, while the coupling between the $O_h$ and $T_d$ units is antiferromagnetic, as illustrated in Fig.~\ref{fig_theo_ferri_alt}. To complement our analysis, we calculated the RIXS-CD in the ferrimagnetic phase using the same models developed above by applying uniform Weiss fields within each $O_h$ and $T_d$ unit, but staggered between them, with the amplitude set to 3~meV (the same value as used in the altermagnetic state above) to simulate the ferrimagnetic state. In the ferrimagnetic state (Fig.~\ref{fig_theo_ferri_alt}), the RIXS-CD is allowed for geometries with both $m'$ and $m$ scattering planes, whereas the latter does not allow CD in the bi-altermagnetic phase, as demonstrated in the main text. Thus, the selection rules differ in the two magnetic states. In addition, we find that the RIXS-CD in the ferrimagnetic state is significantly stronger than in the bi-altermagnetic state, although its amplitude depends on the magnitude of the Weiss field. This is not surprising, given that the excitation energies are site-specific to the $T_d$ and $O_h$ sites, and the signals from the magnetic sublattices add more constructively in the ferrimagnetic alignment, in contrast to the bi-altermagnetic phase, where the two magnetic sublattices exhibit partial cancellation in the $m'$ geometry.



\clearpage
\section{Spin-Orbit Coupling in the $^5T_{2g}$ Manifold} 

We will determine the relation between the atomic spin-orbit coupling constant $ \zeta $ and the effective spin-orbit coupling constant $ \zeta' $ in the $ ^5T_{2g} $ manifold for a $ d^6 $ ion in the octahedral crystal field, and then discuss the ground state of the $T_{1g}$ manifold.

\subsubsection*{1. Full Spin-Orbit Hamiltonian Expectation Value}

The full atomic spin-orbit Hamiltonian is given by:
\begin{equation}
H_1 = \zeta \sum_i \mathbf{l}_i \cdot \mathbf{s}_i.
\end{equation}
We consider a 6-electron configuration in the $t_{2g}$ and $e_g$ orbitals, representing the high-spin $t^{4}_{2g}e_g^2$ configuration. or the $e_g$ level, both orbitals are filled with spin-up electrons.
For the $ t_{2g}$ level, we fill all three $ t_{2g} $ orbitals with spin-up electrons and one with spin-down:
\begin{equation}
\uparrow_{xy}, \uparrow_{yz}, \uparrow_{zx}, \downarrow_{xy}
\end{equation}
Let us denote the one-electron basis set from the $t_{2g}$ level, which have an effective angular momentum $l_{\rm eff}=1$ and spin $s=\frac{1}{2}$ by
$\ket{m_l, m_s}$, where $m_l = \pm1, 0$ and $m_s = \pm \tfrac{1}{2}$.
The full spin-orbit Hamiltonian is:
\begin{equation}
H_1 = \zeta \sum_{i=1}^4 \mathbf{l}_i \cdot \mathbf{s}_i = \zeta \sum_{i=1}^4 (l_{i,z} s_{i,z} + \tfrac{1}{2}(l_{i,+}s_{i,-} + l_{i,-}s_{i,+}))
\end{equation}
Since the basis states are eigenstates of $l_{z}$ and $s_{z}$, only the $l_{z}s_{z}$ terms contribute in this state; the off-diagonal terms vanish. We then have
\begin{equation}
\langle H_1 \rangle = \zeta \sum_{i=1}^{4} \langle m_l^{(i)} m_s^{(i)} | l_{z}^{(i)} s_{z}^{(i)} | m_l^{(i)} m_s^{(i)} \rangle = \zeta \sum_{i=1}^{4} m_l^{(i)} m_s^{(i)}
\end{equation}
Considering the individual states explicitly:
\begin{align*}
(1)\ & \ket{1,\tfrac{1}{2}} \Rightarrow m_l = +1,\ m_s = +\tfrac{1}{2} \Rightarrow m_l m_s = +\tfrac{1}{2} \\
(2)\ & \ket{0,\tfrac{1}{2}} \Rightarrow m_l = 0,\ m_s = +\tfrac{1}{2} \Rightarrow m_l m_s = 0 \\
(3)\ & \ket{-1,\tfrac{1}{2}} \Rightarrow m_l = -1,\ m_s = +\tfrac{1}{2} \Rightarrow m_l m_s = -\tfrac{1}{2} \\
(4)\ & \ket{-1,-\tfrac{1}{2}} \Rightarrow m_l = -1,\ m_s = -\tfrac{1}{2} \Rightarrow m_l m_s = +\tfrac{1}{2}
\end{align*}
Summing all contributions:
\begin{equation}
\langle H_1 \rangle = \zeta \left( \tfrac{1}{2} + 0 - \tfrac{1}{2} + \tfrac{1}{2} \right) = \zeta \cdot \tfrac{1}{2}
\end{equation}
That is,
\begin{equation}
\langle H_1 \rangle = \frac{\zeta}{2}
\end{equation}

\subsubsection*{2. Effective Hamiltonian in the $ ^5T_{2g} $ Manifold}

In the $ t_{2g} $ manifold, the effective angular momentum is $ L_{\text{eff}} = 1 $, and the effective Hamiltonian becomes:
\begin{equation}
H_2 = \zeta' \mathbf{L}_{\text{eff}} \cdot \mathbf{S}
\label{soc_8}
\end{equation}
If we ignore the contributions of the two $e_g$ electrons in the $ ^5T_{2g} $ term, and use the identity
\begin{equation}
\mathbf{L}_{\text{eff}} \cdot \mathbf{S} = \frac{1}{2} \left[ J_{\text{eff}}(J_{\text{eff}}+1) - L_{\text{eff}}(L_{\text{eff}}+1) - S(S+1) \right],
\label{soc_9}
\end{equation}
the spin-orbit coupling (SOC) is
\begin{equation}
\mathbf{L}_{\text{eff}} \cdot \mathbf{S} = \frac{1}{2}(12 - 2 - 6) = 2
\Rightarrow \langle H_1 \rangle = 2\zeta',
\end{equation}
for $S = 2$, $L_{\text{eff}} = 1$ and $J = \lvert{L+S}\rvert = 3$.


\vspace{2mm}
{\noindent}We now equate the two results and have
\begin{equation}
\langle H_1 \rangle = \langle H_2 \rangle
\Rightarrow \frac{\zeta}{2} = 2\zeta'
\Rightarrow \zeta' = \frac{\zeta}{4}. 
\end{equation}

{\noindent}The SOC splits the $^5T_2$ state into triplet, quintet, and septet states with effective angular momentum $J_{\rm eff} = 1$, 2, and 3. Using Eqs. \ref{soc_8} and \ref{soc_9}, one can obtain the energy splittings between $J_{\rm eff} = 1$ and 2, and between 2 and 3; they are $\tfrac{1}{2}\zeta$ and $\tfrac{3}{4}\zeta$, respectively.

\subsubsection*{3. Spin-Orbit-Coupled Ground State}

The lowest spin-orbital state $T_{1g}$ of the $^5T_{2g}$ manifold can also be regarded as an $J_{\rm eff} = 1$ state and its eigenfunctions $|J,J_z\rangle$, where $z$ is taken parallel to the crystallographic $c$ axis of Fe$_2$Mo$_3$O$_8$, are given by: 
\begin{gather}
 |J_{\rm eff}=1,J_{{\rm eff},z}=1\rangle = \sqrt\frac{1}{10}|1,0\rangle\rangle-\sqrt\frac{3}{10}|0,1\rangle\rangle+\sqrt\frac{3}{5}|-1,2\rangle\rangle,\notag \\
 |J_{\rm eff}=1,J_{{\rm eff},z}=0\rangle = \sqrt\frac{3}{10}|1,-1\rangle\rangle-\sqrt\frac{2}{5}|0,0\rangle\rangle+\sqrt\frac{3}{10}|-1,1\rangle\rangle, \\
 |J_{\rm eff}=1,J_{{\rm eff},z}=-1\rangle = \sqrt\frac{3}{5}|1,-2\rangle\rangle-\sqrt\frac{3}{10}|0,-1\rangle\rangle+\sqrt\frac{1}{10}|-1,0\rangle\rangle,\notag
\end{gather}
where $|L_{{\rm eff},z},S_z\rangle\rangle \equiv |L_{{\rm eff},z}\rangle |S_z\rangle $.
The $z$-components of the magnetic moments are $-\tfrac{5}{2}\mu_B$, $0$, and $\tfrac{5}{2}\mu_B$, respectively. 
The magnetic moment is given by $\boldsymbol{\mu} = -g {\mu_B}\mathbf{J}$. For $J_{\rm eff}=1$, $L_{\rm eff}=1$, and $S=2$, the Landé $g$-factor is $g = \tfrac{5}{2}$. 
Therefore, the $z$-components of the magnetic moment for $J_z = 1$, 0, and $-1$ are $-\tfrac{5}{2}\mu_B$, $0$, and $\tfrac{5}{2}\mu_B$, respectively.
The magnitude of the magnetic moment $\tfrac{5}{2}\mu_B$ is smaller than the $S=2$ spin-only value of 4 $\mu_B$ due to the anti-parallel contributions of the orbital magnetic moment.

\bibliographystyle{unsrt}
\bibliography{reference}